\documentclass[10pt]{article}

\usepackage[normalem]{ulem}
\usepackage[utf8]{inputenc}
\usepackage[toc,page]{appendix}

\usepackage{amsmath,amsfonts,amssymb,amsthm}
\usepackage{graphicx}
\usepackage{float}
\usepackage{hyperref}
\usepackage[format=plain,font=small,textfont=it]{caption}
\usepackage[dvipsnames]{xcolor}
\usepackage{mathrsfs}
\usepackage{physics}
\usepackage{empheq}
\usepackage{dsfont} 
\usepackage{algorithm} 
\usepackage{enumitem}

\usepackage{subcaption}

\usepackage{listings}

\graphicspath{{pic/}}

\newcommand{\sltc}{{$SL(2,\mathbb{C})$}}

\newcommand{\Wsix}[6]{\left \{ \begin{array}{ccc} #1 & #2 & #3 \\ #4 & #5 & #6 \end{array}\right \} }

\newcommand{\dref}[2]{#1(\subref{#2})}

\begin{document}

\title{Radiative corrections to the Lorentzian EPRL spin foam propagator}

\author{\Large{Pietro Don\`a${}^{a}$\footnote{dona.pietro@gmail.com}, \ } \Large{Pietropaolo Frisoni${}^{b}$\footnote{pfrisoni@uwo.ca}, \ } \Large{Edward Wilson-Ewing${}^c$\footnote{edward.wilson-ewing@unb.ca} \ }
\smallskip \\ 
\small{\textit{${}^a$ Center for Space, Time and the Quantum, 13288 Marseille, France}}\\
\small{\textit{${}^b$ Department of Physics and Astronomy, University of Western Ontario, London, ON N6A 5B7, Canada}} \\
\small{\textit{${}^c$ Department of Mathematics and Statistics, University of New Brunswick, Fredericton, NB E3B 5A3, Canada}}
}

\date{}

\maketitle

\begin{abstract}
\noindent We numerically estimate the divergence of several two-vertex diagrams that contribute to the radiative corrections for the Lorentzian EPRL spin foam propagator. We compute the amplitudes as functions of a homogeneous cutoff over the bulk quantum numbers, fixed boundary data, and different Immirzi parameters, and find that for a class of two-vertex diagrams, those with fewer than six internal faces are convergent. The calculations are done with the numerical framework \texttt{sl2cfoam-next}.
\end{abstract}


\section{Introduction}
\label{sec:intro}

The main goal of spin foam theory is to define the dynamics of loop quantum gravity in a background independent and Lorentz covariant way, providing transition amplitudes between spin network states \cite{Rovelli2015, Perez:2012wv}. The state of the art are the EPRL and FK spin foam models \cite{Engle:2007wy, Freidel:2007py}; in this paper we will focus on the Lorentzian EPRL spin foam model. These theories have a compelling connection with discrete general relativity in the double limit of finer discretization and vanishing $\hbar$ \cite{Barrett:2009mw, Dona:2020yao, Dona:2020tvv, Han:2021kll, Engle:2021xfs}. 

\medskip

The theory is ultraviolet finite; however, the unbounded summation over the bulk degrees of freedom can cause large-volume infrared divergences (although the infrared divergences are not present in the extension of the theory that includes a cosmological constant using quantum groups \cite{Han:2011aa}). A complete spin foam theory requires a renormalization procedure to remove these low-energy divergences, and renormalization will be essential to define the continuum limit properly. Various renormalization procedures have been proposed in the context of 2-complex refinement \cite{Dittrich:2014mxa, Bahr:2015gxa}, or in group field theory \cite{Bonzom:2011br, Baratin:2013rja, Finocchiaro:2020fhl}, but despite this work there remain many important open questions.

An important step in developing a renormalization procedure in spin foam models is to consider `self-energy' corrections to the propagator. These radiative corrections to the EPRL model have been studied analytically \cite{Riello:2013bzw}, numerically \cite{Frisoni:2021uwx,Frisoni:2021dlk}, and using hybrid techniques \cite{Dona:2018pxq}. So far, the main object of study has been a particular Feynman diagram believed to give the leading-order radiative correction; this Feynman diagram (which is often called the `melonic' self-energy diagram) is a particular two-vertex spin foam diagram associated with a space-time bubble. To calculate the contribution from this diagram, the common strategy is to introduce a homogeneous cutoff on all the spin foam bulk face summations and estimate the degree of divergence of the amplitude, studying it as a function of the cutoff. There are strong numerical indications that the divergence for the melonic self-energy graph is linear in the cutoff \cite{Frisoni:2021uwx,Frisoni:2021dlk}. There are also some general (although weaker) bounds on the degree of divergence for any spin foam diagram \cite{Mikovic:2011bh}.

In general, different spin foam amplitudes are associated with triangulations sharing the same fixed boundary. We order the spin foam diagrams having the same boundary by the number of vertices. The melonic self-energy diagram is not the only two vertex diagrams contributing to the radiative corrections to the Lorentzian EPRL propagator, but most of the other diagrams have a smaller number of bulk faces. Therefore, we expect most of them to contribute to the self-energy calculation with subdominant divergences. This is indeed the case in topological $SU(2)$ BF theory, where the calculation can be done analytically (see Appendix~\ref{app:SU2}), but even if there are some clear structural analogies between the two theories (in particular, they share the same propagator), there remain some important differences and an explicit computation of the divergence of all diagrams is needed for the EPRL spin foam model.

\medskip

In this paper, we continue the numerical study of the two-vertex diagrams contributing to the Lorentzian EPRL propagator started in \cite{Frisoni:2021uwx}. We focus on the subclass of diagrams with two vertices and two boundary edges that are each contained by the four boundary faces. The diagrams differ in the connectivity of the internal edges and faces, and in particular the number of internal faces varies from one diagram to another. Analytical calculations of transition amplitudes with the Lorentzian EPRL spin foam theory are challenging and limited to the large spin regime. Recently, there has been significant progress in numerical techniques for spin foam models, and we use these new developments for our numerical analysis. Specifically, we use the numerical techniques from \cite{Frisoni:2021uwx}, including the approximation scheme clarified in \cite{Dona:2022dxs}. The numerical calculations are done using the library \texttt{sl2cfoam-next} \cite{Gozzini:2021kbt} (available in the repository \cite{sl2cfoam_next}), the state of the art code to compute EPRL spin foam amplitudes on a computer.

\medskip

In Section~\ref{sec:diagrams} we review the Lorentzian EPRL transition amplitude and point out the origin of infrared divergences, and we also list a class of diagrams contributing to the two-vertex correction to the propagator. Then, in Section~\ref{sec:numerics} we overview the main ingredients of our numerical recipe. We conclude with Section~\ref{sec:results} where we analyze the amplitudes and estimate the divergence of the spin foam diagrams we study. For completeness, we also include a detailed expression of the Lorentzian EPRL vertex amplitude in Appendix~\ref{app:vertexdetails}, the analytic calculation of the divergent two-vertex diagrams for the topological $SU(2)$ BF theory in Appendix~\ref{app:SU2}, and a discussion concerning other self-energy diagrams (that should be included or not in the calculation depending on the symmetries of the spin foam model) in Appendix~\ref{app:other}.


\section{EPRL radiative corrections}
\label{sec:diagrams}

Spin foam diagrams are constructed by contracting interaction vertices with spin foam propagators along the spin foam edges. The Lorentzian EPRL spin foam model prescribes a simplicial%
\footnote{A general form of the spin foam vertex exists for an arbitrary number of edges \cite{Kaminski:2009fm}, but we do not consider it here since in this case there is an infinite number of diagrams contributing to the radiative corrections to the theory. We restrict to the simplicial vertex to avoid this uncontrolled proliferation of diagrams.}
interaction vertex and the associated vertex amplitude is 
\begin{equation}
\label{eq:vertexamplitude}
    A_v \left(j_f, \  i_e\right) = \raisebox{-0.4\height}{\includegraphics[scale=1]{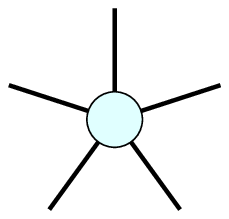}} =   \sum_{l_f=j_f}^\infty \sum_{k_e} \{ 15 j\} (j_f,l_f)  \prod_{e = 2}^{5}  B_4^\gamma \left( l_f, j_f ; i_{e} ,k_{e} \right) (2k_{e}+1)  \ ,
\end{equation}
The amplitude \eqref{eq:vertexamplitude} has fifteen arguments (ten spins $j_f$ and five intertwiners $i_e$), we report its detailed definition in Appendix~\ref{app:vertexdetails}. The form of the amplitude in \eqref{eq:vertexamplitude} is known as the booster decomposition of the vertex amplitude and rewrites the vertex amplitude as a superposition of ${15j}$ symbols weighted by booster functions $B_4^\gamma$ \cite{Speziale:2016axj}. The booster functions enforce the simplicity constraints of the EPRL spin foam models, depend on the Immirzi parameter $\gamma$, and have a compelling geometrical interpretation in terms of boosted tetrahedra \cite{Dona:2020xzv}. We refer to the original paper \cite{Speziale:2016axj} for an explicit expression in terms of intertwiners and \sltc \ $\gamma$-simple unitary irreducible representations. The EPRL propagator is simple and is given by 
\begin{equation}
    \label{eq:edgeamplitude}
    A_e \left(i_e\right) = \raisebox{-0.5\height}{\includegraphics[scale=1]{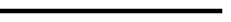}} = (2 i_e +1)\  \delta_{i_e, i_{e}'}  \ ,
\end{equation}
forcing the intertwiners in two vertices corresponding to the same edge to be the same. 

\medskip

A spin foam transition amplitude associated with the 2-complex of a triangulation $\Delta$ is given by the contraction of vertices and propagators dual to the 2-complex, spins and intertwiners associated with bulk faces and edges are summed over. Each face is weighted with a dimensional factor $A_f(j_f) = 2j_f+1$ and the whole amplitude is
\begin{equation}
\label{eq:transitionamplitude}
    A_{\Delta} = \sum_{j_f, i_e}  \prod_f A_f(j_f) \prod_e A_e(i_e) \prod_v A_v \left(j_f, \  i_e\right) \ .
\end{equation}

\medskip

\begin{figure}[H]
\centering
    \begin{subfigure}{0.45\textwidth}
        \centering
        \includegraphics[scale=1]{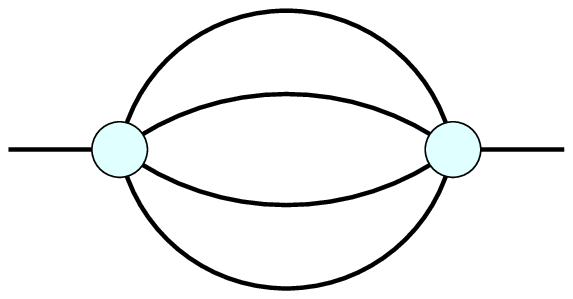}
        \caption{}
        \label{fig:selfenergy}
    \end{subfigure}
\hfill
    \begin{subfigure}{0.45\textwidth}
        \centering
        \includegraphics[scale=1]{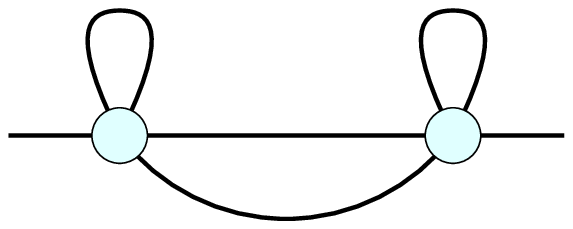}
        \caption{}
        \label{fig:frog}
    \end{subfigure}
\hfill
    \begin{subfigure}{0.45\textwidth}
        \centering
        \includegraphics[scale=1]{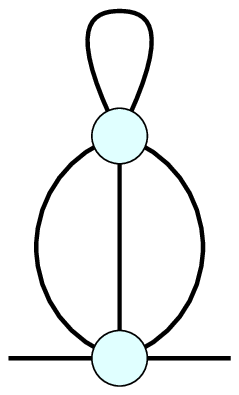}
        \caption{}
        \label{fig:penguin}
    \end{subfigure}
\hfill
    \begin{subfigure}{0.45\textwidth}
        \centering
        \includegraphics[scale=1]{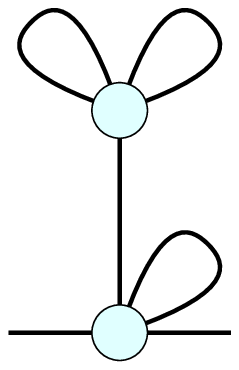}
        \caption{}
        \label{fig:goat}
    \end{subfigure}
\caption{Classes of spin foam diagrams with two vertices and two boundary edges.}
\label{fig:alldiagrams}
\end{figure}

We focus on diagrams contributing radiative corrections that satisfy three requirements to avoid an excessive proliferation of diagrams to study. Each diagram we consider has
\begin{enumerate}
    \item Two vertices
    \item Two boundary edges
    \item Four boundary faces, and each one of these faces contains both boundary edges.
    \item Only trivial propagators, preserving the order of the faces.
\end{enumerate}
We give some examples of diagrams excluded by these conditions in Appendix~\ref{app:other}. There are four classes of diagrams that differ by the combinatorics of the bulk edges shared by the two vertices, as shown in Figure~\ref{fig:alldiagrams}. 

The diagrams in each class differ by the combinatorics of the faces in the bulk. For the diagrams satisfying the 4 conditions listed above, we find: (a) two diagrams with 6 unbounded faces, one in class \ref{fig:selfenergy} and one in class \ref{fig:goat}, (b) in each of class \subref{fig:frog} and \subref{fig:goat} there are two diagrams with four unbounded faces, and there is another one in class \subref{fig:penguin}, (c) finally, there are two diagrams in class \subref{fig:selfenergy} with two unbounded faces. To complete the analysis, we also mention that there are other diagrams across all four classes where the sums on the internal spins and intertwiners are fully bounded due to $SU(2)$ triangular inequalities, in which case the corresponding amplitude is a sum of a limited number of terms and the summation is trivially finite.

\section{Numerical calculations}
\label{sec:numerics}

We study the divergences of the Lorentzian EPRL spin foam diagrams of Figure~\ref{fig:alldiagrams} using \texttt{sl2cfoam-next}. The library is an open-source library written in \texttt{C} to calculate EPRL transition amplitudes numerically, with an optional user-friendly \texttt{Julia} interface. It is based on the booster decomposition of the vertex amplitude \eqref{eq:vertexamplitude} and implements the calculation of the vertex amplitude with a homogeneous truncation of the unbounded sums over the virtual spins $l_f$ parameterized by $\Delta l$
\begin{equation}
    \sum_{l_f=j_f}^\infty \to \sum_{l_f=j_f}^{j_f + \Delta l} \ .
\end{equation}

We compute the diagrams following the strategy described in \cite{Frisoni:2021uwx, Frisoni:2021dlk} where the diagram in class \subref{fig:selfenergy} with 6 internal faces is studied in great detail. We refer to those articles for an updated analysis of the divergence of the melonic self-energy EPRL spin foam diagram.

We fix the boundary spins $j_b$ and intertwiners $i_b$ and choose a truncation parameter $\Delta l$. The sums over the spin of the bulk faces $j_f$ are unbounded and potentially divergent. We introduce a homogeneous cutoff $K$ on these sums while we sum over all the possible values of the bulk intertwiners $i_e$. We take this chance to stress the difference between $\Delta l$ and $K$. At first glance, they look similar, as they are both introduced as an upper bound of a previously unbounded summation but are profoundly different. The parameter $\Delta l$ represents the unavoidable truncation to approximate the \emph{convergent} series in the vertex amplitude. On the other hand, $K$ is the cutoff on the divergent sums on the bulk degrees of freedom responsible for the large volume divergences of the diagrams. 

The parameters $\Delta l$ and $K$ are independent. Nevertheless, we have some technical constraints. We want a value for the cutoff $K$ large enough to estimate the degree of divergence of the amplitude but compatible with the computational resources at our disposal. We also want a truncation $\Delta l$ large enough to obtain a good numerical approximation of the amplitude. Increasing the truncation will improve the numerical estimate of the amplitude; however, although the resources necessary for the calculation will grow considerably with the truncation, the numerical accuracy will improve very little. It is essential to choose a balanced value of the parameters. Previous work suggests that choosing similar values for $\Delta l$ and $K$ gives good numerical results \cite{Frisoni:2021uwx}. Given the computational resources at our disposal we choose $\Delta l = 10$, and consider $K$ running from 0 to 10 in half-integer steps.

In the following, we neglect the dimensional factors related to boundary faces and intertwiners, as these correspond to a constant multiplicative factor that does not affect the functional dependence of the amplitude on $K$.

To minimize the dependence on the parameter $\Delta l$, we use the convergence acceleration technique described in \cite{Dona:2022dxs}. Using the value of the amplitude at fixed $K$ obtained with the three largest truncations available $\Delta l$, $\Delta l-1$, and $\Delta l-2$, we approximate the amplitude (corresponding to the limit $\Delta l \to \infty$) with
\begin{equation}
  \label{eq:acceleration}
 A(K) \approx \frac{A(K,\Delta l) A(K, \Delta l-2) - A^2(K,\Delta l-1)}{A(K, \Delta l) - 2A(K, \Delta l-1)+A(K, \Delta l-2)} \ .
\end{equation}
The \texttt{Julia} scripts used to compute the diagrams, the Wolfram's Mathematica notebooks in which we analyze the amplitudes, and the numerical values of the amplitudes are available in the public repository \cite{Radiative_repository}. The scripts can be parallelized on multiple CPU cores to optimize the numerical calculations. For each spin foam diagram, the sums over the spins of the bulk faces shared by the two vertex amplitudes \eqref{eq:vertexamplitude} are distributed on the available tasks, using the distributed memory parallel computing implementation of \texttt{Julia}. Then, the sums over the spins of the bulk faces on every single vertex are parallelized on the threads of each task. Finally, the contraction over bulk intertwiners uses the loop vectorization technique. The code we used is scalable on a supercomputer. We provide the script used for the calculations on the Compute Canada clusters in the repository \cite{Radiative_repository}. The total CPU usage for the computations described in this paper consisted in $\sim$ 200 CPU hours.

%
%

\section{Results}
\label{sec:results}

We focus on uniform boundary configurations to simplify the numerical calculation and optimize the computational resources at our disposal. We take all the boundary spins $j_b$ to be the same, fixing them to the lowest possible non-trivial value $j_b=\tfrac{1}{2}$. We look at the amplitudes with equal boundary intertwiners and with different Immirzi parameters to minimize the dependence on the boundary data choices we have to make. We compute the amplitudes using a cutoff ranging from $K=0$ to $10$ in half-integer steps and a truncation $\Delta l = 10$. We approximate the amplitudes using \eqref{eq:acceleration}.


\subsection{Spin foam diagrams with 6 bulk faces}

\begin{figure}
\centering
    \begin{subfigure}{0.49\textwidth}
        \centering
        \includegraphics[scale=0.3]{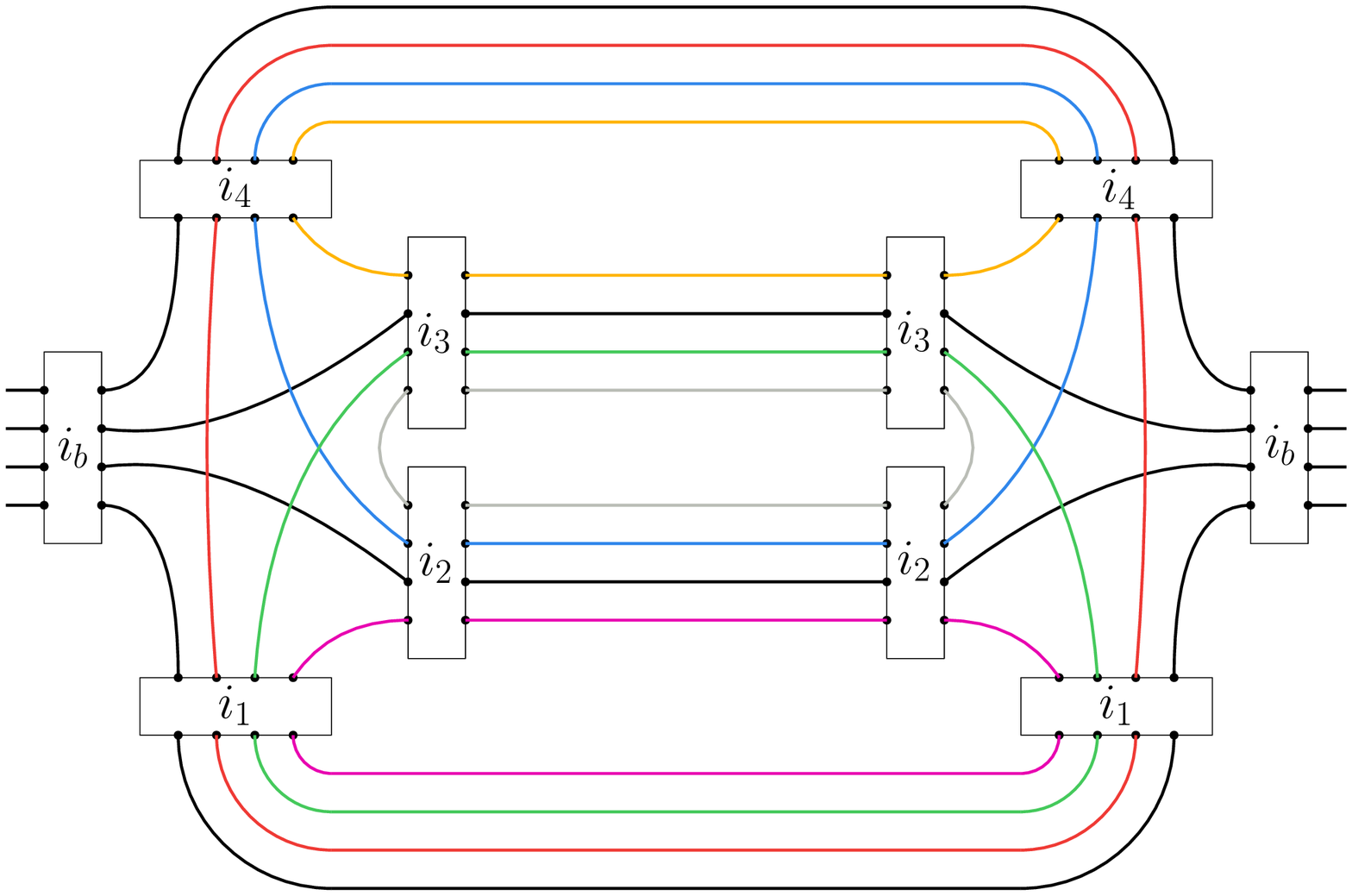}
        \caption{}
        \label{fig:6Fselfenergy}
    \end{subfigure}
\hfill
    \begin{subfigure}{0.49\textwidth}
        \centering
        \includegraphics[scale=0.3]{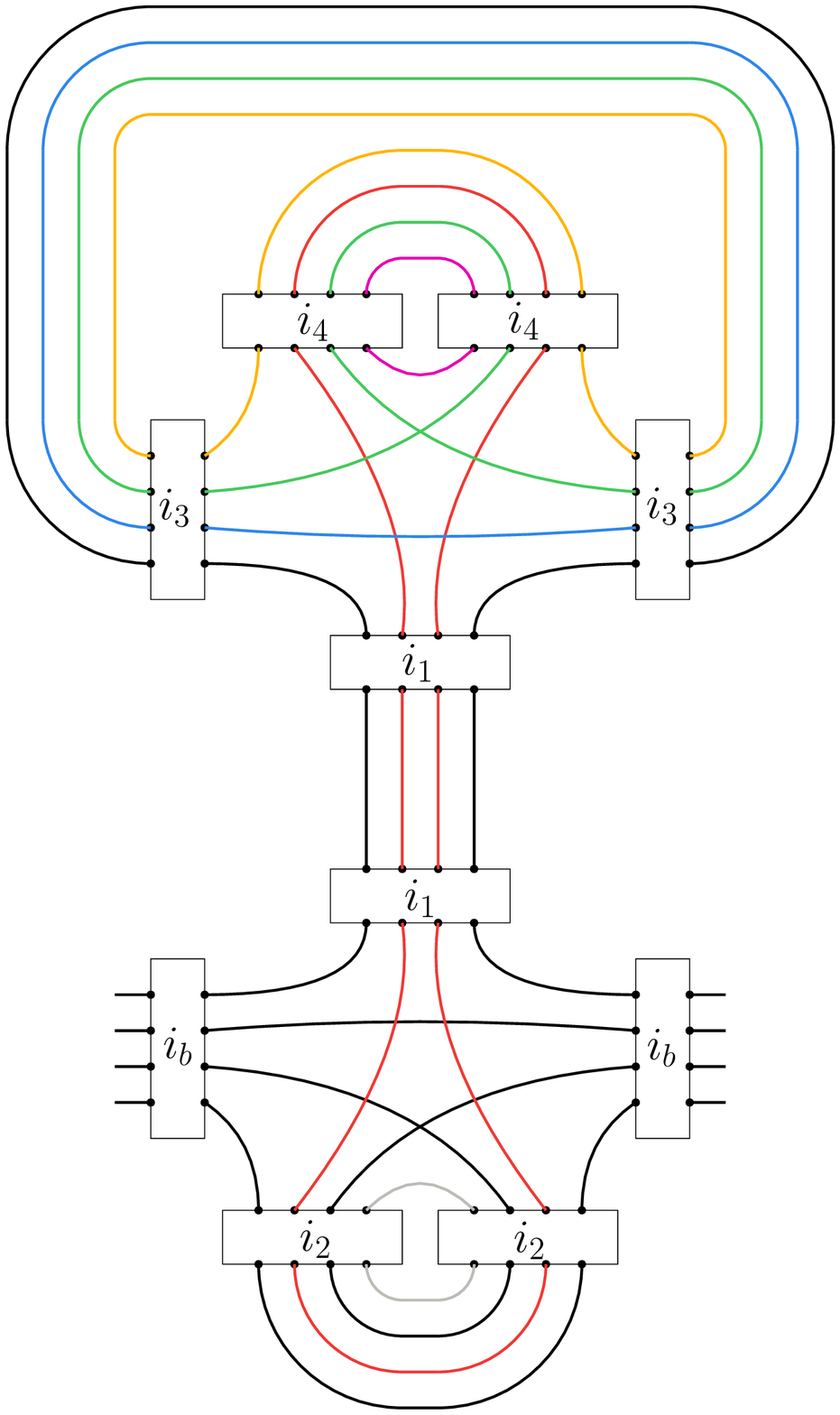}
        \caption{}
        \label{fig:6Fgoat}
    \end{subfigure}
\hfill
\caption{6F. Wiring of the spin foam diagrams with 6 bulk faces. We highlight the internal faces in different colors.}
\label{fig:6faces}
\end{figure}

We start by briefly discussing the two diagrams with the most unbounded bulk faces. We show in Figure~\ref{fig:6faces} the wiring diagrams highlighting the combinatorics of their internal faces.

Studying the divergence of the melonic spin foam diagram \dref{6F}{fig:6Fselfenergy} required tremendous numerical effort, for a detailed presentation of the results see \cite{Frisoni:2021uwx, Frisoni:2021dlk}. Similarly, we expect that the numerical analysis of the spin foam diagram \dref{6F}{fig:6Fgoat} will also require optimization and significant computational time, and for this reason we leave a detailed study for future work. Nonetheless, it is useful to point out a few key points concerning these diagrams before moving on to other diagrams with fewer internal faces.

The EPRL amplitude associated with the diagram \dref{6F}{fig:6Fselfenergy} is
\begin{equation}
    \label{eq:6Fselfenergy}
    \begin{split}
     A_{6F(A)} \left(j_b, \  i_b; \, K\right) = \sum_{j_f = 0}^{K}  \sum_{i_e} & A_f(j_{f_1})A_f(j_{f_2})A_f(j_{f_3})A_f(j_{f_4}) A_f(j_{f_5}) A_f(j_{f_6}) \, A_e(i_{1})A_e(i_{2})A_e(i_{3})A_e(i_{4}) \times\\ 
     & A_v \left(  j_b, j_b, j_b, j_b, j_{f_1}, j_{f_2}, j_{f_3}, j_{f_4}, j_{f_5}, j_{f_6}; \ i_b , i_1, i_2, i_3, i_4\right) \times \\
     &  A_v \left(  j_b, j_b, j_b, j_b, j_{f_1}, j_{f_2}, j_{f_3}, j_{f_4}, j_{f_5}, j_{f_6}; \ i_b , i_4, i_3, i_2, i_1\right) \ .
    \end{split}
\end{equation}

In \cite{Frisoni:2021uwx}, it is shown that the amplitude \eqref{eq:6Fselfenergy} with boundary intertwiners $i_b=0$, boundary spins $j_b=\tfrac{1}{2}$ and Immirzi parameter $\gamma=0.1$ diverges linearly in the cutoff $K$%
\footnote{It is worth mentioning that the result looks slightly different for larger values of $\gamma$. However, there are solid arguments to explain why it is a numerical artifact, and that the divergence of the melonic self-energy diagram is linear in the cutoff.}.
This result is obtained by fitting the amplitude as a function of the homogeneous cutoff up to $K=10$ with truncation $\Delta l = 20$ and using a convergence acceleration technique like \eqref{eq:acceleration}. Crucially, one of the observations of \cite{Frisoni:2021uwx} is that the value of the amplitude estimated using \eqref{eq:acceleration} with truncation $\Delta l = 20$ and $\Delta l = 10$ are essentially identical. This justifies our choice of using a truncation $\Delta l =10$ for the other diagrams to calculate the amplitude accurately while limiting the need for time-intensive computational resources. 

Next, the EPRL amplitude associated with the diagram \dref{6F}{fig:6Fgoat} is
\begin{equation}
    \label{eq:6Fgoat}
    \begin{split}
     A_{6F(B)} \left(j_b, \  i_b; \, K\right) = \sum_{j_f = 0}^{K}  \sum_{i_e} & A_f(j_{f_1})A_f(j_{f_2})A_f(j_{f_3})A_f(j_{f_4}) A_f(j_{f_5}) A_f(j_{f_6}) \, A_e(i_{1})A_e(i_{2})A_e(i_{3})A_e(i_{4}) \times\\ 
     & A_v \left(  j_b, j_{f_1}, j_{f_1}, j_b, j_b, j_b, j_b, j_{f_2}, j_b, j_b; \ i_1 , i_b, i_2, i_2, i_b\right) \times \\
     &  A_v \left(  j_b, j_{f_1}, j_{f_1}, j_b, j_{f_3}, j_{f_4}, j_{f_5}, j_{f_6}, j_{f_4}, j_{f_3}; \ i_1 , i_3, i_4, i_4, i_3\right) \ .
    \end{split}
\end{equation}
The number of terms of the amplitude \dref{6F}{fig:6Fgoat} for increasing values of the cutoff $K$ grows approximately as $ \sim 30 \cdot K^4 $. Although this trend is slower than it is for the melonic self-energy diagram \cite{Frisoni:2021uwx}, there is an important difference that makes the computation of amplitude  \dref{6F}{fig:6Fgoat} far more resource-demanding.
In the melonic self-energy diagram \dref{6F}{fig:6Fselfenergy}, there is a boundary face on each edge, and triangular inequalities constrain the growth of the spin of the bulk faces at each of the two vertices. Note that this does not happen in the diagram \dref{6F}{fig:6Fgoat}, and as a result we expect that the calculation of the vertices requires considerable computational resources. As $K$ increases, eight spins out of ten in one vertex of \dref{6F}{fig:6Fgoat} are of the order of $K$, compared to the six in the diagram \dref{6F}{fig:6Fselfenergy}. For the same reason, the computation time as a function of the truncation parameter $\Delta l$ is also considerably higher. As a rough estimate, the calculation of the amplitude \dref{6F}{fig:6Fgoat} up to $K = 10$ and $\Delta l = 2$, distributed on 32 CPUs, took about 16 hours. The calculation with $\Delta l =10$ could take weeks. The calculation of the amplitude \dref{6F}{fig:6Fgoat} with sufficient precision requires an enormous investment of time and computational resources that we do not currently have access to, and therefore we leave a detailed study of this diagram for future work.


\subsection{Spin foam diagrams with 4 bulk faces}

\begin{figure}
\centering
    \begin{subfigure}{0.49\textwidth}
        \centering
        \includegraphics[scale=0.3]{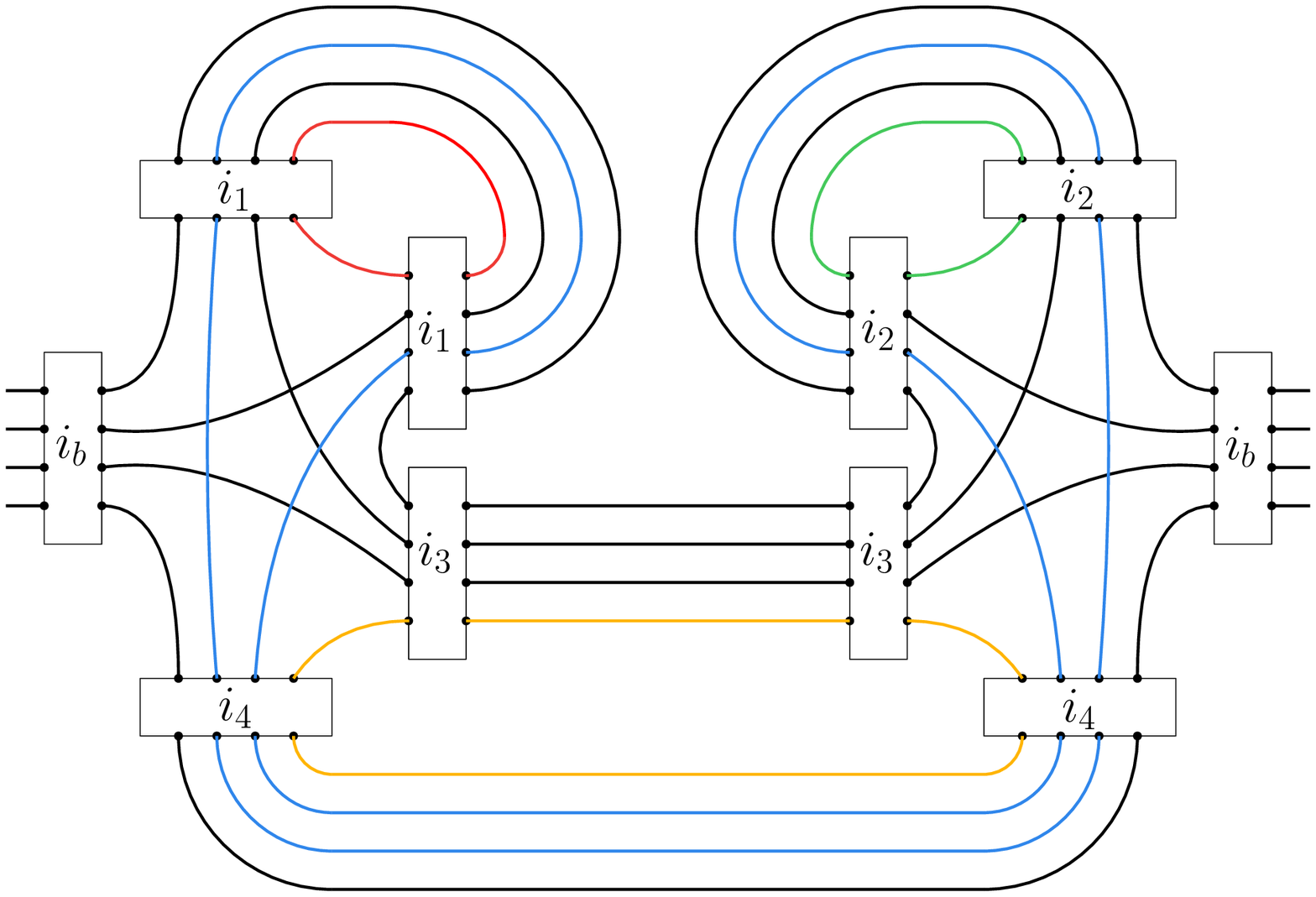}
        \caption{}
        \label{fig:4Ffrog1}
    \end{subfigure}
\hfill
    \begin{subfigure}{0.49\textwidth}
        \centering
        \includegraphics[scale=0.3]{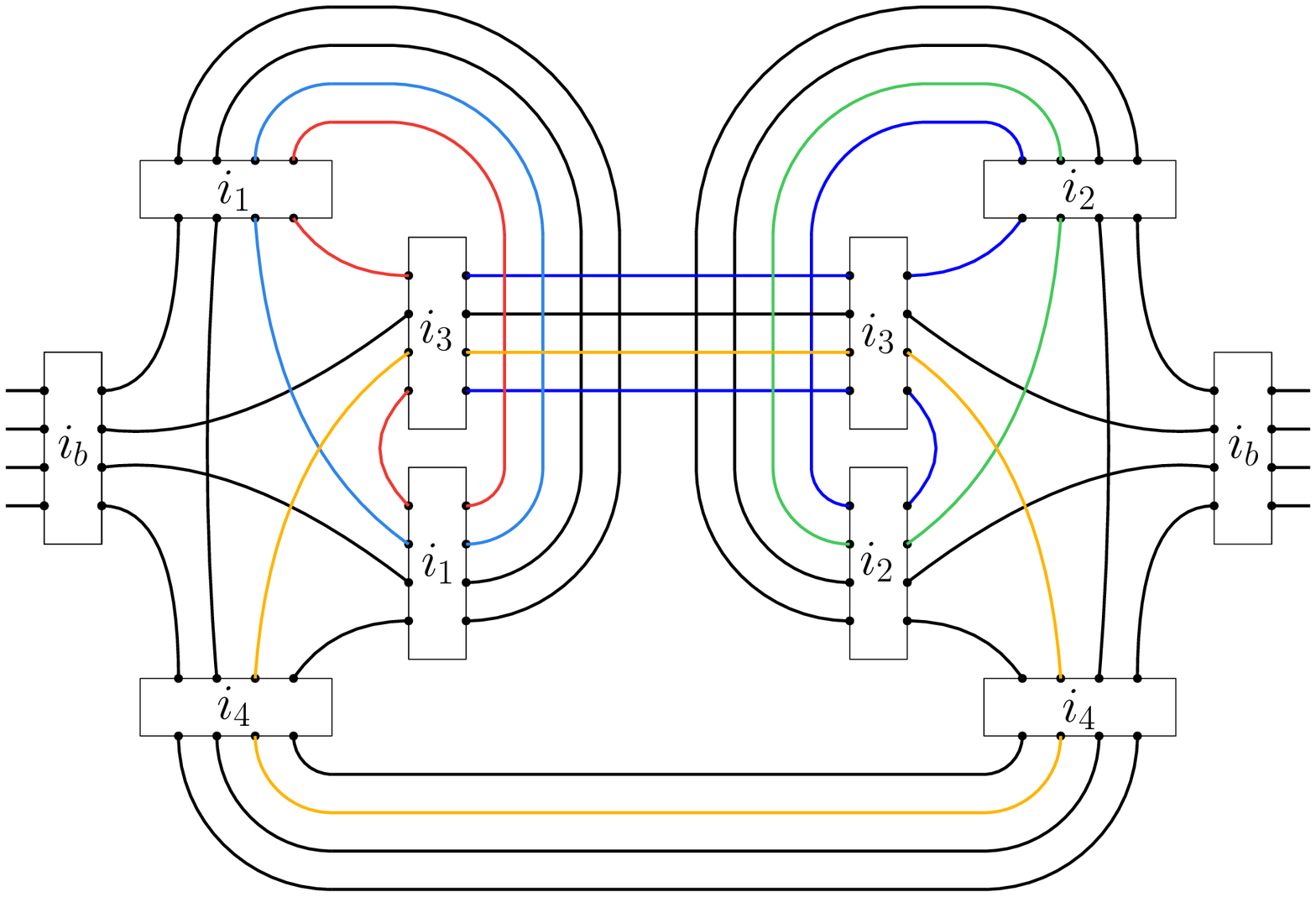}
        \caption{}
        \label{fig:4Ffrog2}
    \end{subfigure}
\hfill
    \begin{subfigure}{0.3\textwidth}
        \centering
        \includegraphics[scale=0.3]{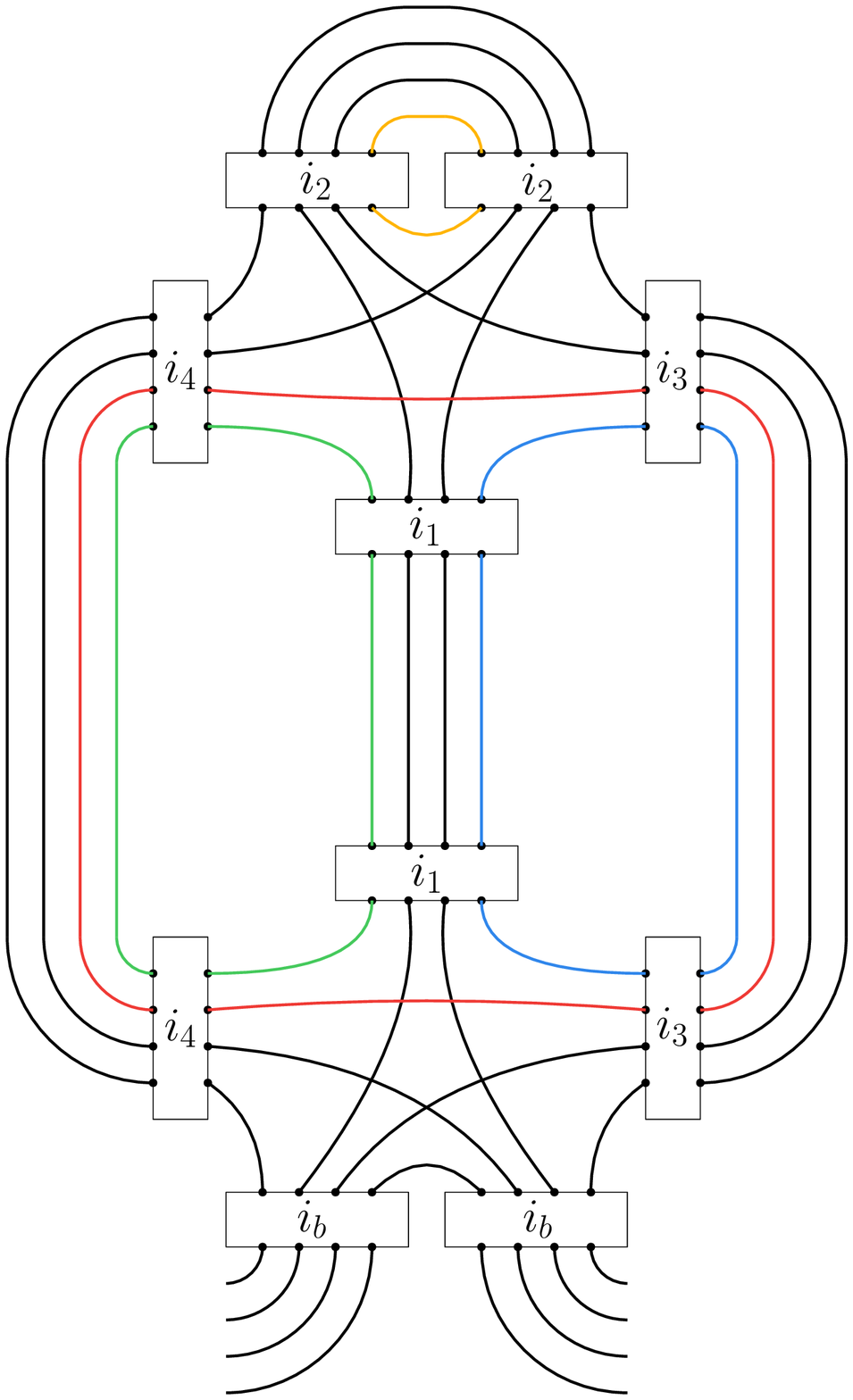}
        \caption{}
        \label{fig:4Fpenguin}
    \end{subfigure}
\hfill
    \begin{subfigure}{0.3\textwidth}
        \centering
        \includegraphics[scale=0.3]{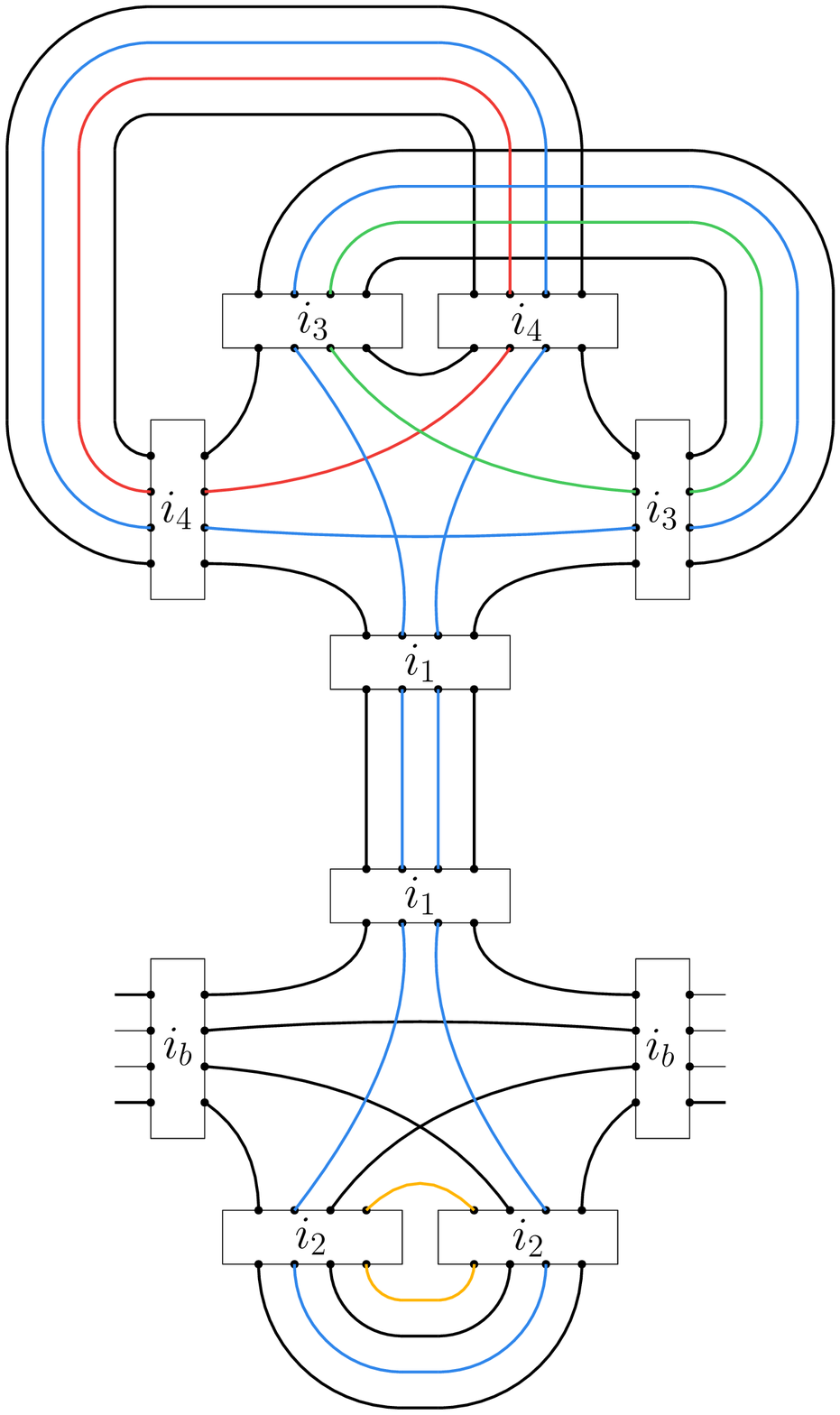}
        \caption{}
        \label{fig:4Fgoat2}
    \end{subfigure}
\hfill
    \begin{subfigure}{0.3\textwidth}
        \centering
        \includegraphics[scale=0.28]{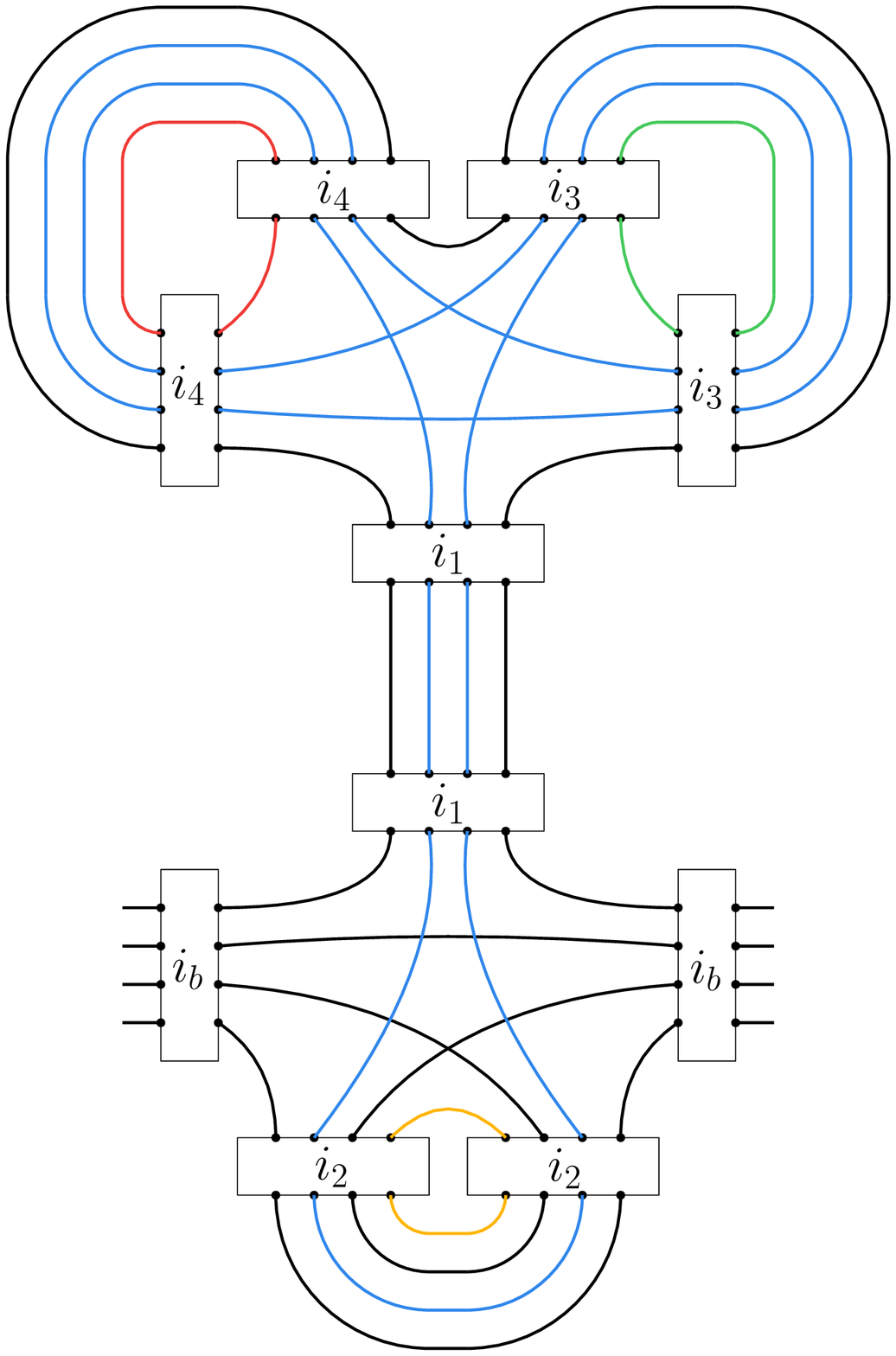}
        \caption{}
        \label{fig:4Fgoat1}
    \end{subfigure}
\caption{4F. Wiring of the spin foam diagrams with 4 bulk faces. We highlight the internal faces in different colors.}
\label{fig:4faces}
\end{figure}

There are five diagrams (satisfying the four conditions given above) with four unbounded bulk faces. Their wiring diagrams are given in Figure~\ref{fig:4faces}, showing the combinatorics of the internal faces.

\begin{figure}
\centering
    \includegraphics[scale=0.9]{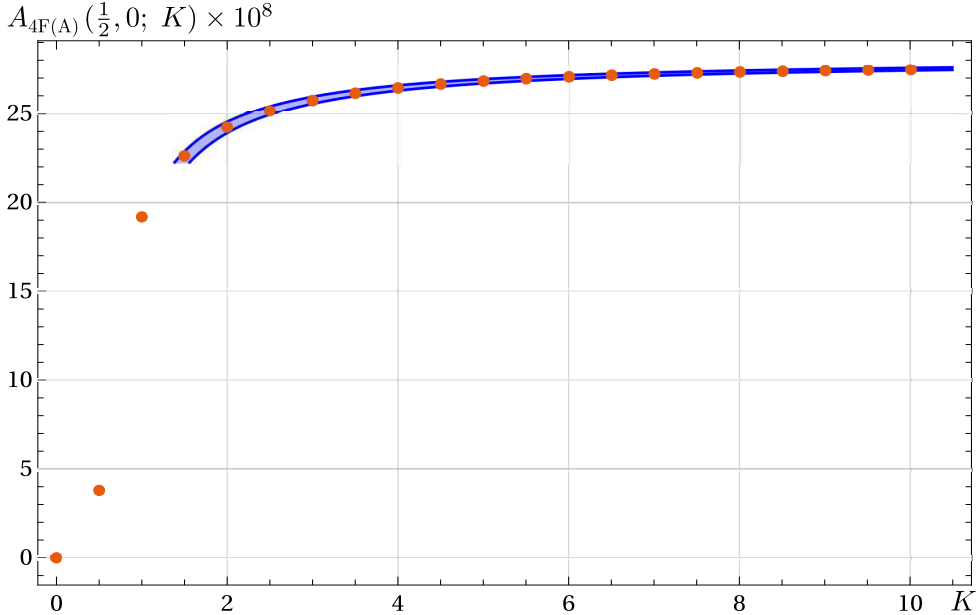}
    \caption{\label{fig:frogplot} Plot of the numerical values of the EPRL amplitude of the spin foam \eqref{fig:frog} with boundary spins $j_b=\tfrac{1}{2}$, boundary intertwiners $i_b=0$ and Immirzi parameter $\gamma=0.1$. We plot with a blue band the fit \eqref{eq:frogfinal}.}
\end{figure}

We illustrate the numerical analysis in detail for one diagram; the others are very similar, and we simply report the results for the other diagrams. The EPRL amplitude associated with the spin foam diagram \dref{4F}{fig:4Ffrog1} is 
\begin{equation}
\label{eq:frogamplitude1}
    \begin{split}
     A_{4F(A)} \left(j_b, \  i_b; \, K\right) = \sum_{j_f = 0}^{K} \sum_{i_e} & A_f(j_{f_1})A_f(j_{f_2})A_f(j_{f_3})A_f(j_{f_4}) \, A_e(i_{1})A_e(i_{2})A_e(i_{3})A_e(i_{4}) \times\\ 
     &  A_v \left( j_b, j_b, j_b, j_b, j_{f_1}, j_{f_2}, j_{f_2}, j_{b}, j_{b},  j_{f_3} ; \ i_b , i_4, i_3, i_1, i_1\right) \times \\
     & A_v \left( j_b, j_b, j_b, j_b, j_{f_4}, j_{b},  j_{f_2}, j_{b}, j_{f_2}, j_{f_1} ; \ i_b , i_2, i_2, i_3, i_4\right) \ .
    \end{split}
\end{equation}
We start by looking at the amplitude with both boundary intertwiners $i_b=0$. We report the numerical values of the amplitude for different cutoffs $K$ in Figure~\ref{fig:frogplot}.

In topological BF theory, the amplitude of this diagram diverges less rapidly than the melonic self-energy diagram (see Appendix~\ref{app:SU2}), and it seems reasonable to expect that the same will be true for the EPRL spin foam model. Since there is strong numerical evidence that the EPRL melonic self-energy diagram is linearly divergent in the cutoff, we expect the diagram \dref{4F}{fig:4Ffrog1} will have a degree of divergence that is at most $\log K$.

We make a baseline fit with a four-parameter function using the \texttt{NonlinearModelFit} routine in Mathematica, with the result
\begin{equation}
 A_{4F(A)}^{prelim}(\tfrac{1}{2},0; K) =  - 4.11 \cdot 10^{-9} \log K + 2.93 \cdot 10^{-7} -8.61 \cdot 10^{-8} K^{-1} -1.77 \cdot 10^{-8} K^{-2} \ .
\end{equation}
The coefficient of $\log K$ is orders of magnitude smaller than the others, indicating that the amplitude may be convergent. In general, it is difficult to numerically distinguish between a logarithmic and a convergent behavior at these scales, since $K$ is at most 10 in the numerical results. Increasing the cutoff by a factor 2 results in a $\log(2)\approx 0.69$ increase of the logarithmic term but costs several more hours of computational time. The amount of resources needed for the calculation does not scale linearly in the cutoff (a rough estimate of the computational time would be $K$ to the power of the number of unbounded summations). As a result, increasing the cutoff is not a good strategy to answer this question.

There is a more efficient way to verify the convergence of the diagram. We amplify the divergence of the diagram by replacing the face amplitude in \eqref{eq:transitionamplitude}
\begin{equation}
    A_f(j_f) = 2j_f + 1 \quad \longrightarrow \quad A_f^{amp}(j_f) = (2j_f + 1)^2 \ .
\end{equation}
We can verify that of the four unbounded sums in \eqref{eq:frogamplitude1}, only three are really unbounded since $SU(2)$ triangular inequalities limit the spin on one of the internal faces (specifically, the yellow internal face which shares an edge with three fixed boundary faces). Therefore, the degree of divergence of the amplified amplitude increases by $K^3$. We evaluate the amplified amplitude numerically using the same parameters and boundary data as the original one. We fit the amplified amplitude with a cubic polynomial in the cutoff and find that the coefficient of $K^3$ is negligible by two orders of magnitude with respect to the others%
\footnote{Fitting with a cubic polynomial, we find a ratio between the coefficients of $K^3$ and $K^2$ of $0.01$, indicating that the coefficient of $K^3$ is irrelevant.}. 
The amplified amplitude diverges quadratically as $K^2$ in the cutoff, confirming that the original amplitude had a negative degree of divergence. To avoid any doubt, we also repeat the calculation with an amplified face amplitude $A_f^{amp}(j_f) = (2j_f + 1)^3$, with the expected result that the amplitude is again amplified by a factor of $K^3$, now diverging as the fifth power $K^5$ of the cutoff, confirming that the original amplitude \eqref{eq:frogamplitude1} is convergent%
\footnote{Fitting with a degree six polynomial, we find a ratio of $0.05$ between the coefficients of $K^6$ and $K^5$, suggesting that the coefficient of $K^6$ is irrelevant.}.

Given the strong indications that the leading order dependence on the cutoff scales as $K^{-1}$, we fit the initial amplitude again, omitting the $\log K$ term, and find with a $95\%$ confidence interval
\begin{equation}
\label{eq:frogfinal}
 A_{4F(A)}(\tfrac{1}{2},0; K)  = (2.799, 2.806) \cdot 10^{-7} - (5.387,4.790)\cdot 10^{-8} \ K^{-1} - (5.552,4.501) \cdot 10^{-8} \ K^{-2} \ . 
\end{equation}
To remove the cutoff, it is easy to take the limit $K\to\infty$ with the result $A_{4F(A)} = (2.799, 2.806) \cdot 10^{-7}$.

\medskip 

\begin{figure}
\centering
        \includegraphics[width=0.495\textwidth]{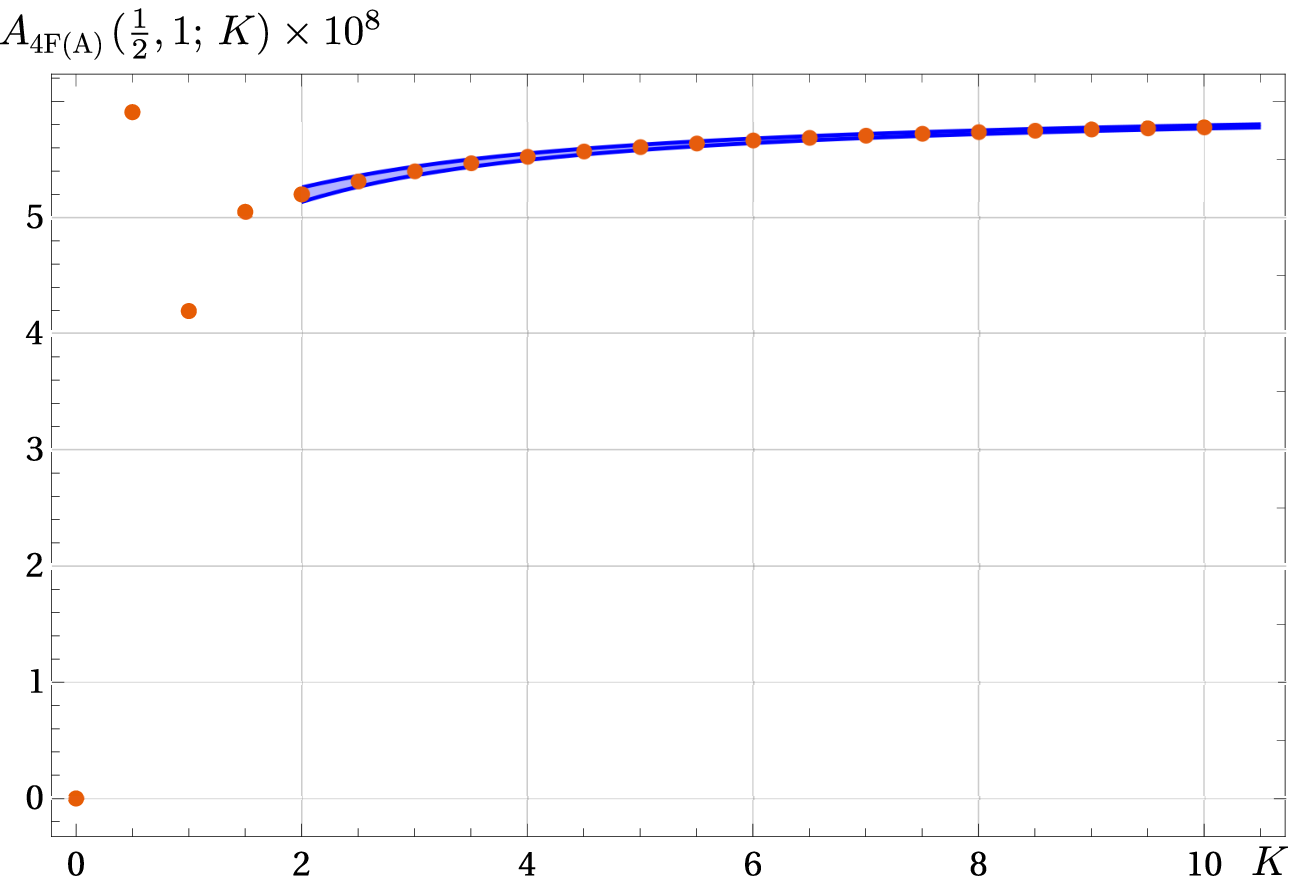}
        \includegraphics[width=0.495\textwidth]{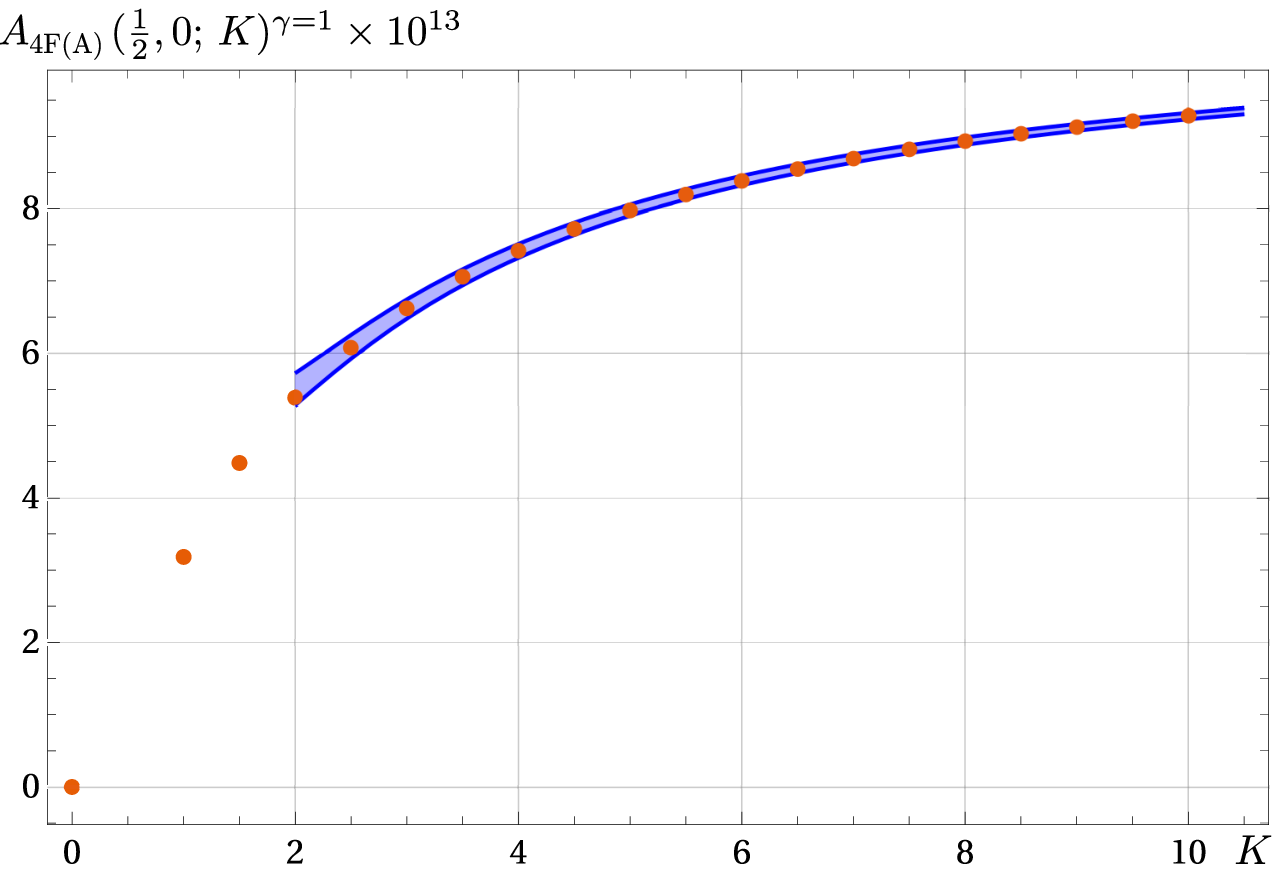}
\caption{\label{fig:frogplotother} Plot of the numerical values of the EPRL amplitude of the spin foam diagram \dref{4F}{fig:4Ffrog1} with boundary spins $j_b=\tfrac{1}{2}$ (orange dots) and the corresponding fit (blue band) \eqref{eq:frogfinalib1} and \eqref{eq:frogfinalig1}. Left panel: boundary intertwiners $i_b=1$ and Immirzi parameter $\gamma =0.1$. Right panel: boundary intertwiners $i_b=0$ and Immirzi parameter $\gamma = 1$. The strange behavior of the amplitude with $i_b=1$ and $K=\tfrac{1}{2}$ is a numerical artifact of this diagram, due to the cutoff value being smaller then the intertwiners. }
\end{figure}

The result concerning the degree of divergence of this diagram seems to be robust. We also studied the amplitude for different boundary intertwiners $i_b=1$ and found a similar outcome. We fit the amplitude and see with a $95\%$ confidence interval
\begin{equation}
\label{eq:frogfinalib1}
 A_{4F(A)}(\tfrac{1}{2},1; K)  = (5.969, 5.983) \cdot 10^{-8} - (2.123,2.006)\cdot 10^{-8} \ K^{-1} - (0.910,1.117) \cdot 10^{-8} \ K^{-2} \ . 
\end{equation}
Likewise, we study the amplitude with boundary intertwiners $i_b=0$ and a larger Immirzi parameter $\gamma=1$. Again, we find that the amplitude is convergent,
\begin{equation}
\label{eq:frogfinalig1}
 A_{4F(A)}(\tfrac{1}{2},0; K) ^{\gamma=1} = (1.080, 1.084) \cdot 10^{-12} - (1.680,1.640)\cdot 10^{-12} \ K^{-1} - (1.146,1.229) \cdot 10^{-12} \ K^{-2} \ . 
\end{equation}
We summarize these results in Figure~\ref{fig:frogplotother}. 

\medskip

\begin{figure}
\centering
        \includegraphics[width=0.495\textwidth]{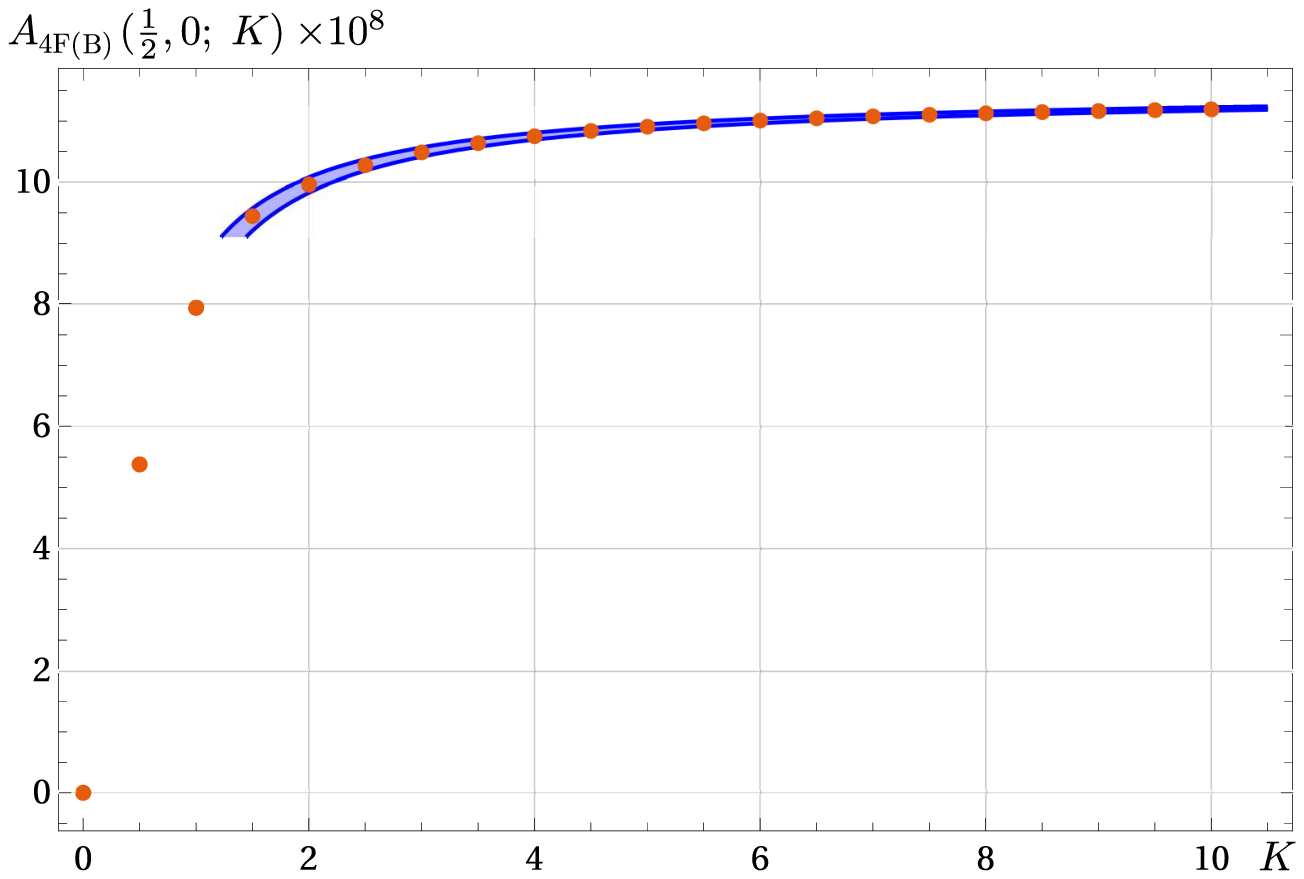}
        \includegraphics[width=0.495\textwidth]{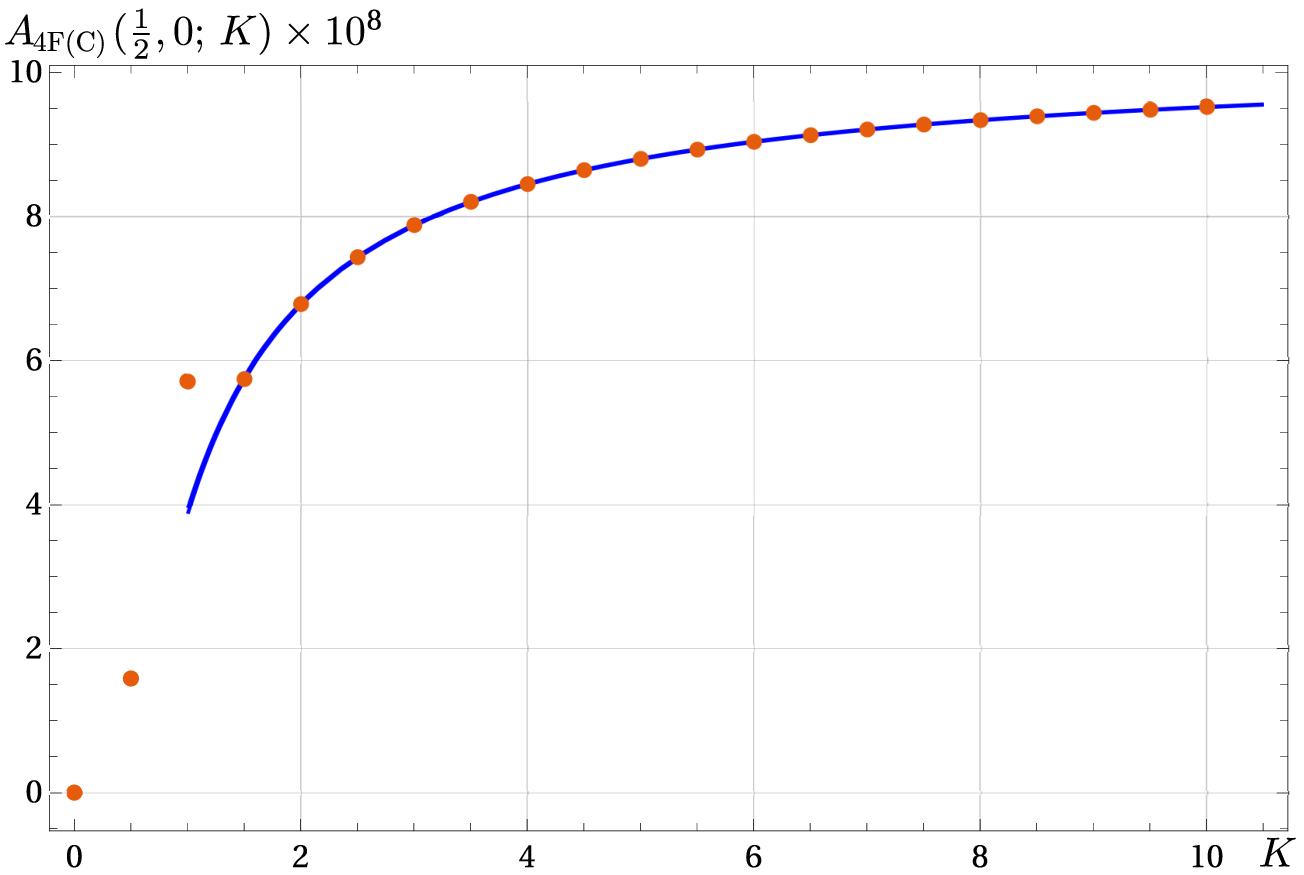}
\caption{\label{fig:frog2andpenguinplot}
Plot of the numerical values of the EPRL amplitudes with boundary spins $j_b=\tfrac{1}{2}$, boundary intertwiners $i_b=0$ and Immirzi parameter $\gamma=0.1$. We plot with a blue band the corresponding fits. 
Left panel: Spin foam diagram \dref{4F}{fig:4Ffrog2} and fit \eqref{eq:frogfinal2}.
Right panel: Spin foam diagram \dref{4F}{fig:4Fpenguin} and fit \eqref{eq:penguinfinal}.
}
\end{figure}

\noindent We perform a similar analysis to calculate the amplitudes of the other diagrams with 4 internal faces in Figure~\ref{fig:4faces}. {In this case, we report the data, fits, and plots only for $\gamma=0.1$ and boundary intertwiners $i_b=0$, to keep the presentation concise, but we repeated the calculation for $\gamma=1$ and $i_b=1$ finding qualitatively similar results.} The amplitude of the spin foam diagram \dref{4F}{fig:4Ffrog2} is
\begin{equation}
    \label{eq:frogamplitude2}
    \begin{split}
     A_{4F(B)} \left(j_b, \  i_b; \, K\right) = \sum_{j_f = 0}^{K}  \sum_{i_e} & A_f(j_{f_1})A_f(j_{f_2})A_f(j_{f_3})A_f(j_{f_4}) \, A_e(i_{1})A_e(i_{2})A_e(i_{3})A_e(i_{4}) \times\\ 
     & A_v \left( j_b, j_b, j_b, j_b, j_b, j_{f_1}, j_b, j_{f_2}, j_{f_3}, j_{f_2} ; \ i_b , i_4, i_1, i_3, i_1\right) \times \\
     & A_v \left( j_b, j_b, j_b, j_b, j_{f_2}, j_{f_4}, j_b, j_{f_2}, j_{f_1}, j_b ; \ i_b, i_2, i_3, i_2 , i_4\right) \ ,
    \end{split}
\end{equation}
and it appears to be convergent. The numerical results are shown in Figure~\ref{fig:frog2andpenguinplot}, together with the best fit curve with a 95\% confidence interval given by
\begin{equation}
\label{eq:frogfinal2}
 A_{4F(B)}(\tfrac{1}{2},0; K)  = (1.147, 1.150) \cdot 10^{-7} - (2.933,2.696)\cdot 10^{-8} \ K^{-1} - (7.095,2.929) \cdot 10^{-9} \ K^{-2} \ . 
\end{equation}

\noindent The amplitude of the spin foam diagram \dref{4F}{fig:4Fpenguin} is
\begin{equation}
    \label{eq:penguinamplitude}
    \begin{split}
     A_{4F(C)} \left(j_b, \  i_b; \, K\right) = \sum_{j_f = 0}^{K}  \sum_{i_e} & A_f(j_{f_1})A_f(j_{f_2})A_f(j_{f_3})A_f(j_{f_4}) \, A_e(i_{1})A_e(i_{2})A_e(i_{3})A_e(i_{4}) \times\\ 
     & A_v \left(  j_{f_1}, j_b, j_b, j_{f_2} , j_b, j_b, j_{f_3},  j_{b}, j_{b},  j_{b} ; \ i_1 , i_4, i_b, i_b, i_3\right) \times \\
     & A_v \left(   j_{f_2}, j_b, j_b,  j_{f_1},   j_{b}, j_{b}, j_{f_3}, j_{f_4}, j_b, j_b ; \ i_1, i_3, i_2, i_2 , i_4\right) \ ,
    \end{split}
\end{equation}
and again we find a clear indication that this amplitude is convergent. We perform a fit and find with a $95\%$ confidence interval
\begin{equation}
\label{eq:penguinfinal}
 A_{4F(C)}(\tfrac{1}{2},0; K)  = (1.0269, 1.0273) \cdot 10^{-7} - (7.595,7.566)\cdot 10^{-8} \ K^{-1} + (1.199,1.250) \cdot 10^{-8} \ K^{-2} \ . 
\end{equation}
The data and the fit are represented in Figure~\ref{fig:frog2andpenguinplot}.

\medskip

\begin{figure}
\centering
        \includegraphics[width=0.499\textwidth]{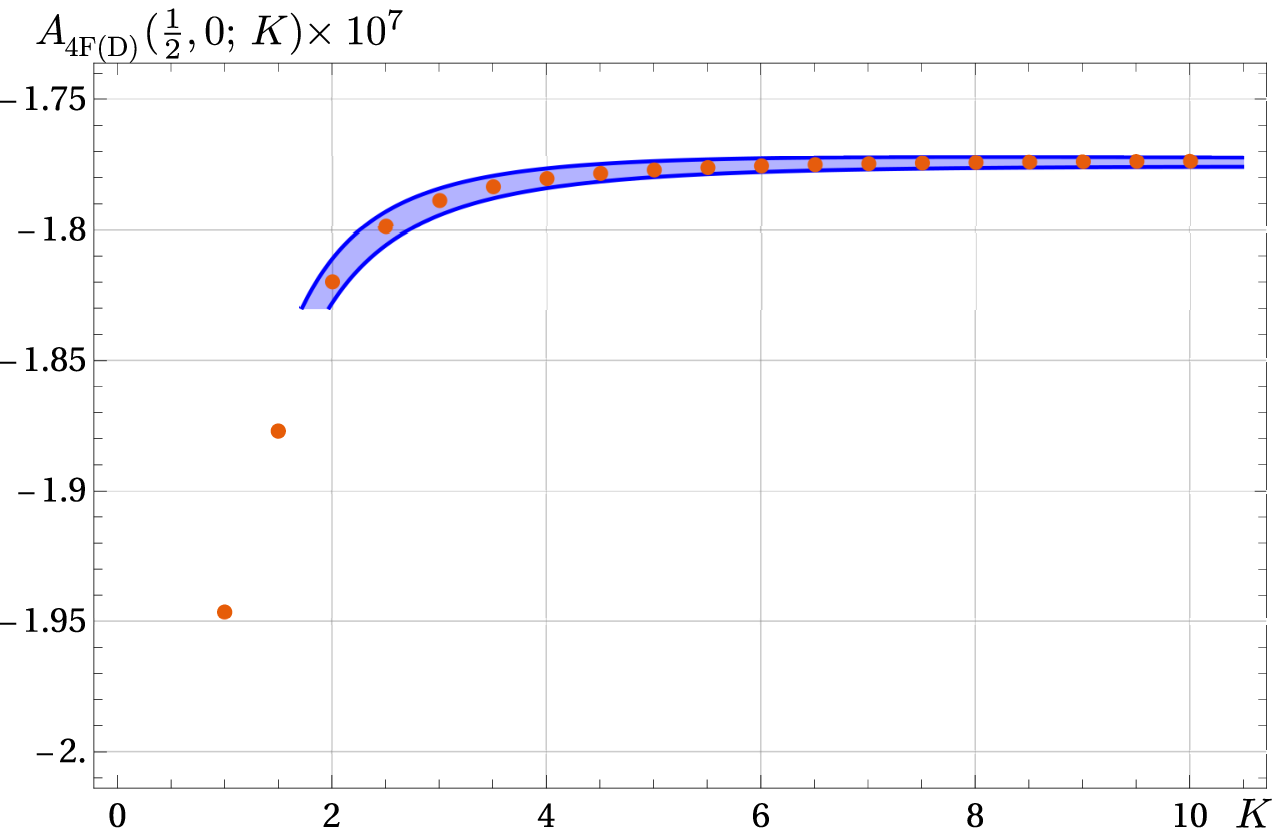}
        \includegraphics[width=0.489\textwidth]{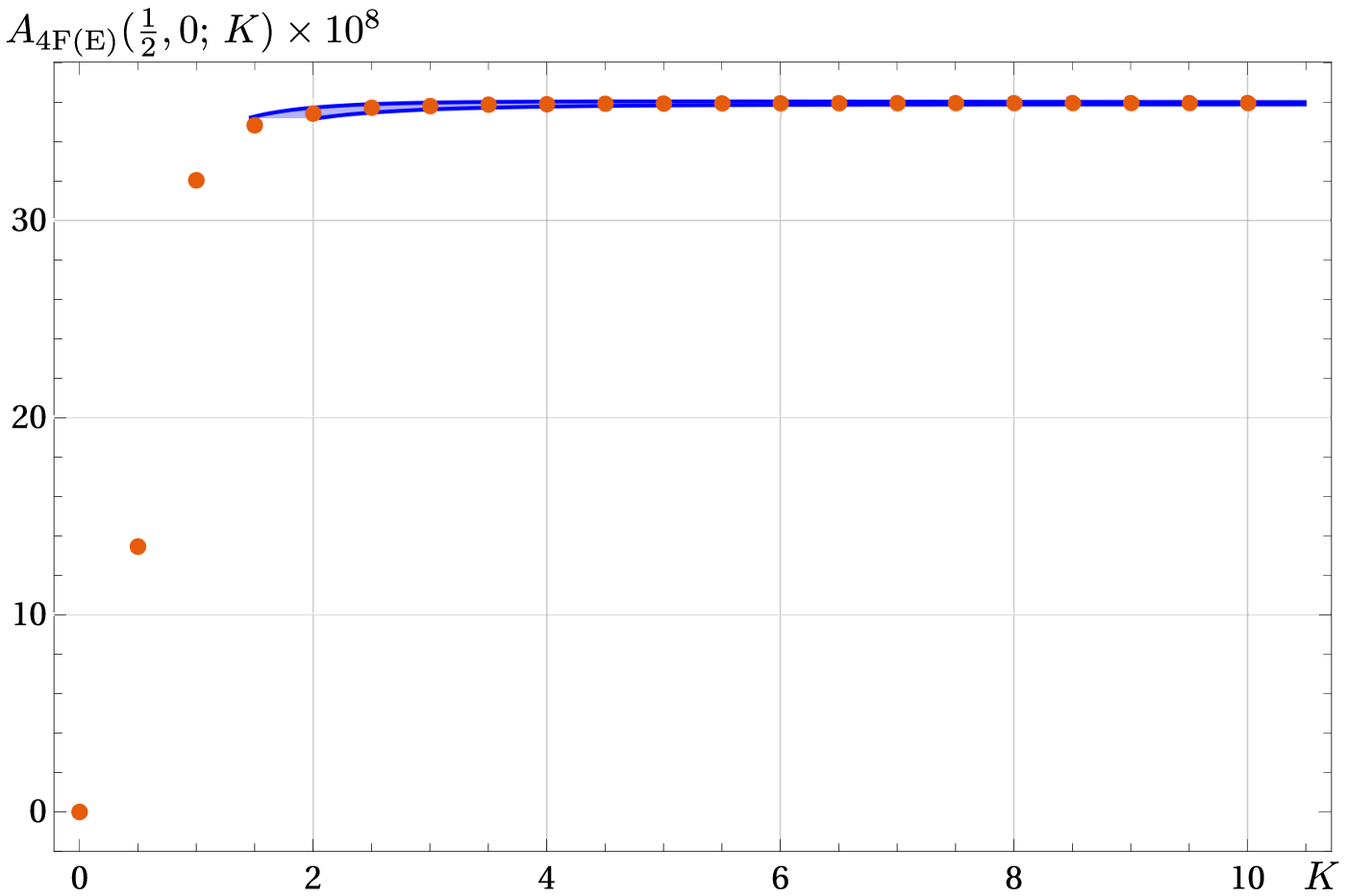}
\caption{
\label{fig:goatplots}
Plot of the numerical values of the EPRL amplitudes with boundary spins $j_b=\tfrac{1}{2}$, boundary intertwiners $i_b=0$ and Immirzi parameter $\gamma=0.1$. We plot with a blue band the corresponding fits. 
Left panel: Spin foam diagram \dref{4F}{fig:4Fgoat2} and fit \eqref{eq:goatfinal}.
Right panel: Spin foam diagram \dref{4F}{fig:4Fgoat1} and fit \eqref{eq:goatfinal2}.
}
\end{figure}

\noindent Finally, we look at the amplitudes \dref{4F}{fig:4Fgoat2} and \dref{4F}{fig:4Fgoat1}. The amplitudes associated with these diagrams are
\begin{equation}
\label{eq:goatamplitude}
    \begin{split}
     A_{4F(D)} \left(j_b, \  i_b; \, K\right) = \sum_{j_f = 0}^{K} \sum_{i_e} &  (-1)^{\chi_D} A_f(j_{f_1})A_f(j_{f_2})A_f(j_{f_3})A_f(j_{f_4}) \, A_e(i_{1})A_e(i_{2})A_e(i_{3})A_e(i_{4}) \times\\ 
     &  A_v \left( j_b, j_{f_1}, j_{f_1}, j_b, j_{b}, j_{b}, j_{b}, j_{f_2}, j_{b},  j_{b} ; \ i_1 , i_b, i_2, i_2, i_b\right) \times \\
     & A_v \left( j_b, j_{f_1}, j_{f_1}, j_b, j_{b}, j_{f_3},  j_{f_1}, j_{b}, j_{f_4}, j_{b} ; \ i_1 , i_3, i_4, i_3, i_4\right) \ , 
    \end{split}
\end{equation}
and 
\begin{equation}
\label{eq:goatamplitude2}
    \begin{split}
     A_{4F(E)} \left(j_b, \  i_b; \, K\right) = \sum_{j_f = 0}^{K} \sum_{i_e} &  (-1)^{\chi_E} A_f(j_{f_1})A_f(j_{f_2})A_f(j_{f_3})A_f(j_{f_4}) \, A_e(i_{1})A_e(i_{2})A_e(i_{3})A_e(i_{4}) \times\\ 
     &  A_v \left( j_b, j_{f_1}, j_{f_1}, j_b, j_b, j_b, j_b, j_{f_2}, j_b, j_b ; \ i_1 , i_b, i_2, i_2, i_b\right) \times \\
     & A_v \left( j_b, j_{f_1}, j_{f_1}, j_b, j_{f_3}, j_{f_1},  j_{f_1}, j_{b}, j_{f_1}, j_{4} ; \ i_1 , i_3, i_3, i_4, i_4\right) \ , 
    \end{split}
\end{equation}
where $\chi_D = 2 j_{f_3}$ and $\chi_E = 2j_b+2j_{f_1}+j_{f_3}+j_{f_4}$ are phases coming from the decomposition of the amplitude. Once again the numerical evaluation of the amplitudes \eqref{eq:goatamplitude} suggests the amplitudes are convergent. We perform a fit with a $95\%$ confidence interval and find 
\begin{equation}
\label{eq:goatfinal}
 A_{4F(D)} (\tfrac{1}{2},0; K)  = -(1.778, 1.776) \cdot 10^{-7} + (5.097,6.681)\cdot 10^{-9} \ K^{-1} - (2.999,2.720) \cdot 10^{-8} \ K^{-2} \ ,
\end{equation}
\begin{equation}
\label{eq:goatfinal2}
 A_{4F(E)} (\tfrac{1}{2},0; K)  = (3.586, 3.592) \cdot 10^{-7} + (8.172,13.486)\cdot 10^{-9} \ K^{-1} - (4.360,3.425) \cdot 10^{-8} \ K^{-2} \ .
\end{equation}
We report these fits and compare them with the data points in Figure~\ref{fig:goatplots}.

It is somewhat surprising that the degree of divergence of these last two diagrams is at most the same as for the first 3 diagrams with four internal faces (and possibly even with a smaller degree of divergence given by $K^{-2}$). Na\"ively, one might expect the degree of divergence to be directly related to the number of unbounded summations. If one looks carefully, diagrams \dref{4F}{fig:4Fgoat2} and \dref{4F}{fig:4Fgoat1} are the only ones among the diagrams in Figure~\ref{fig:4faces} with four unbounded internal faces---the other three diagrams always have one face bounded by triangular inequalities (specifically, the yellow internal face in each diagram). (On the other hand, all of the diagrams in Figure~\ref{fig:4faces} in the $SU(2)$ BF theory have the same degree of divergence.) This counterintuitive result indicates that the degree of divergence depends in a complicated way on the components of the diagram.


\subsection{Spin foam diagrams with 2 bulk faces}

\begin{figure}[H]
\centering
    \begin{subfigure}{0.49\textwidth}
        \centering
        \includegraphics[scale=0.35]{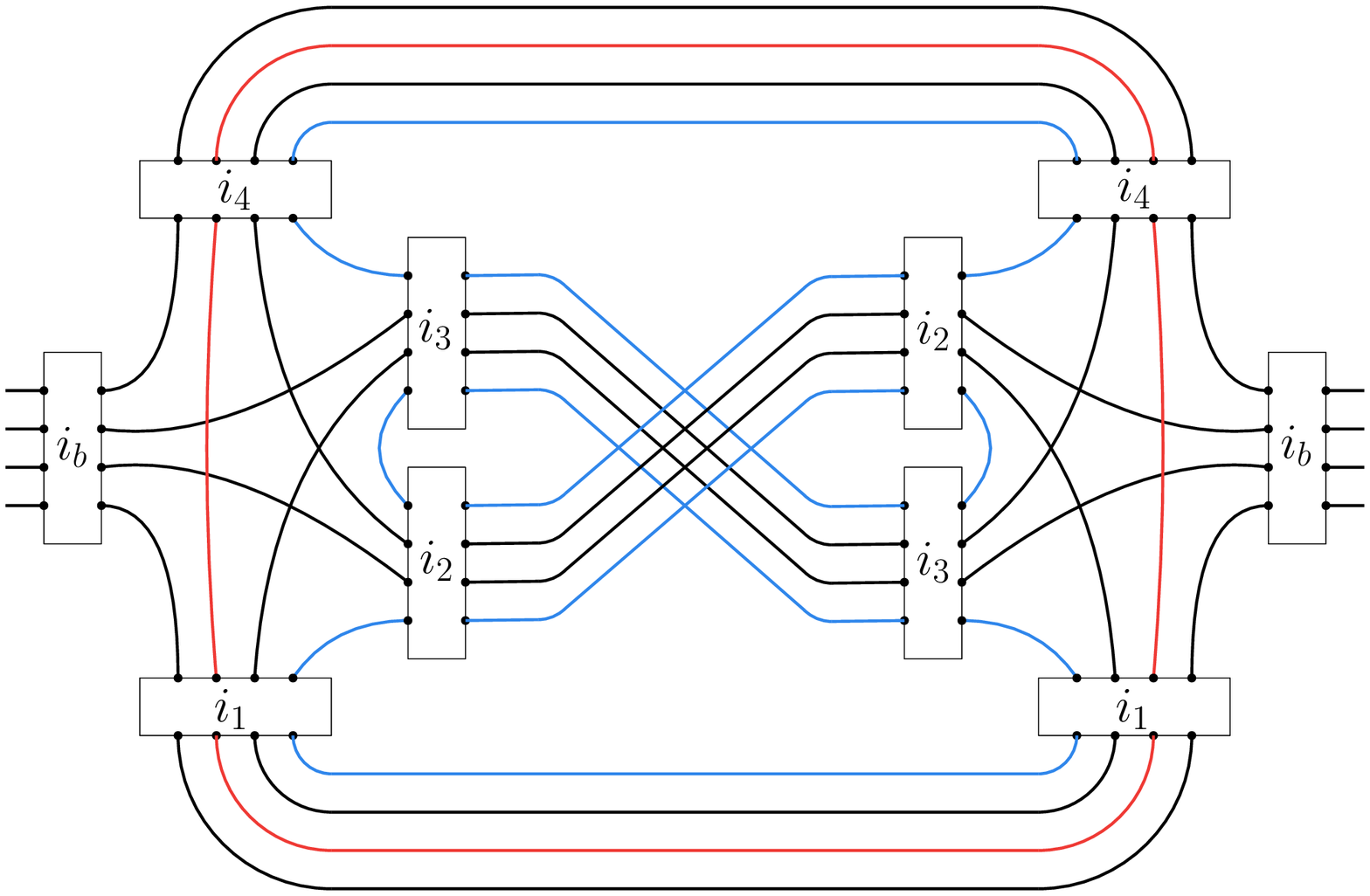}
        \caption{}
        \label{fig:2Fselfenergy1}
    \end{subfigure}
\hfill
    \begin{subfigure}{0.49\textwidth}
        \centering
        \includegraphics[scale=0.3]{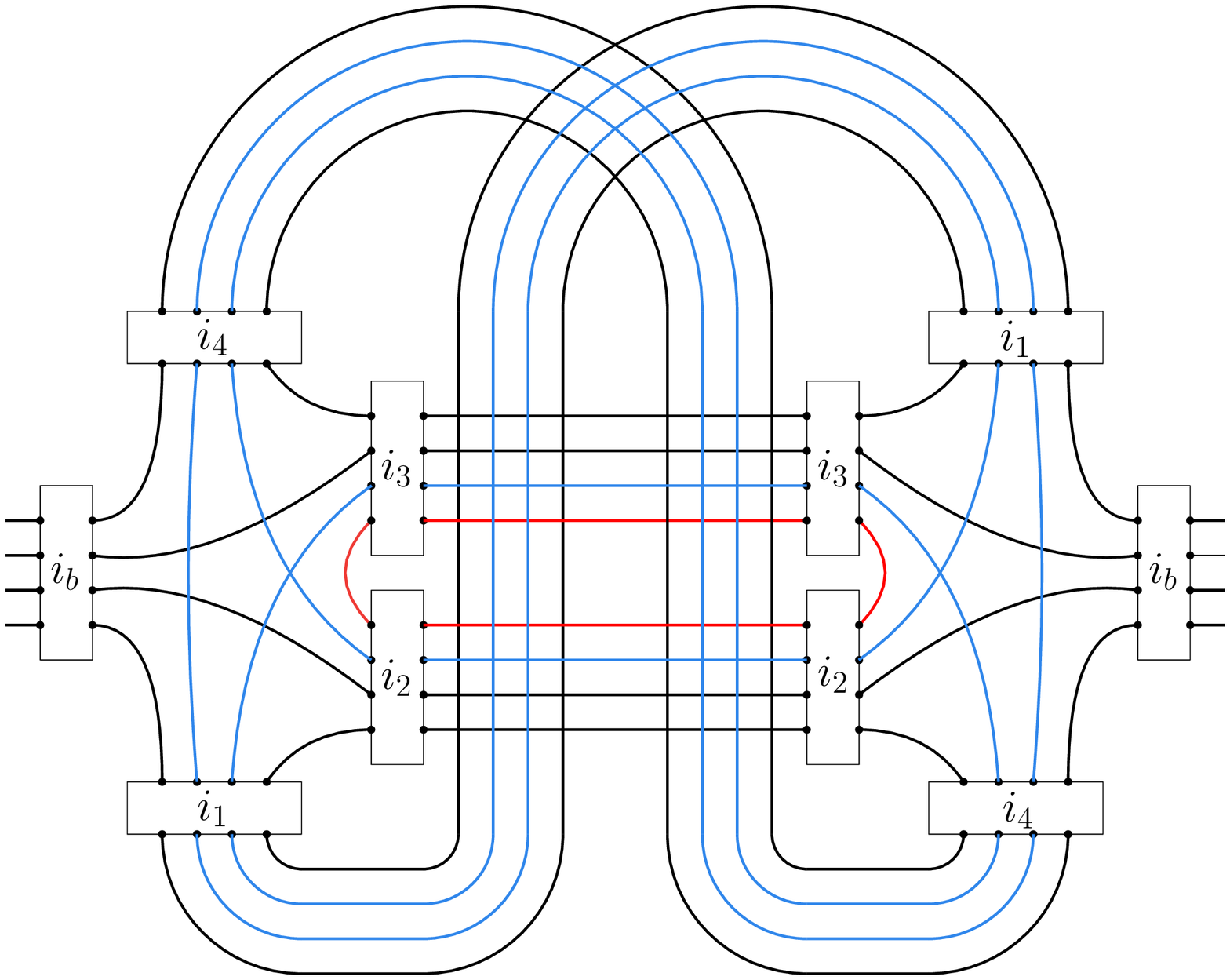}
        \caption{}
        \label{fig:2Fselfenergy2}
    \end{subfigure}
\caption{2F. Wiring of the spin foam diagrams with 2 bulk faces. We highlight the internal faces in different colors.}
\label{fig:2faces}
\end{figure}

There are only two diagrams satisfying the four conditions listed above that have two unbounded bulk faces, these are shown in Figure~\ref{fig:2faces} where we highlight the combinatorics of the internal faces. Note that there exist other diagrams with two internal faces satisfying the four conditions, but in all other such diagrams, the spin of the internal faces is bounded due to $SU(2)$ triangular inequalities. {Also in this case, we report only the calculation with $\gamma=0.1$ and boundary intertwiners $i_b=0$.}

\begin{figure}
\centering
        \includegraphics[width=0.495\textwidth]{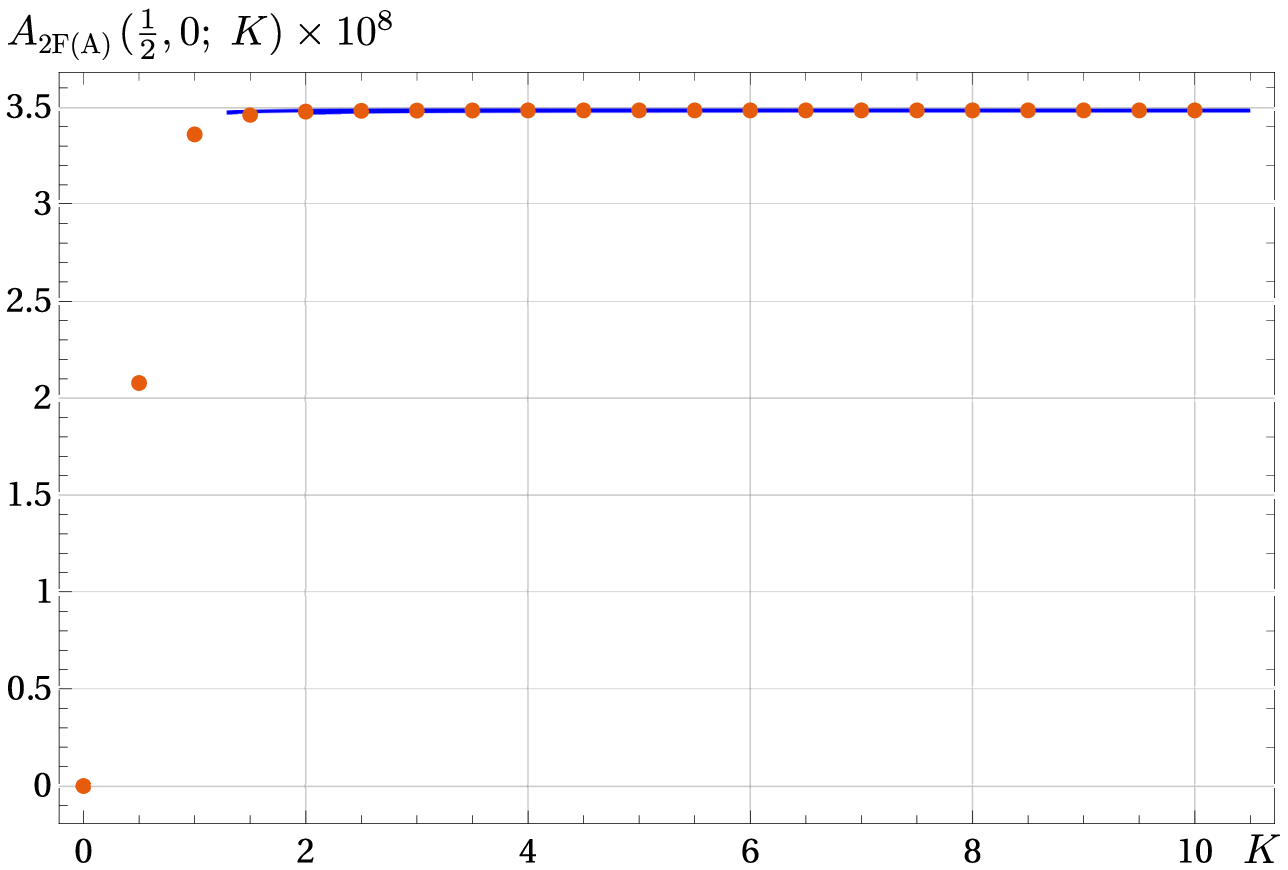}
        \includegraphics[width=0.495\textwidth]{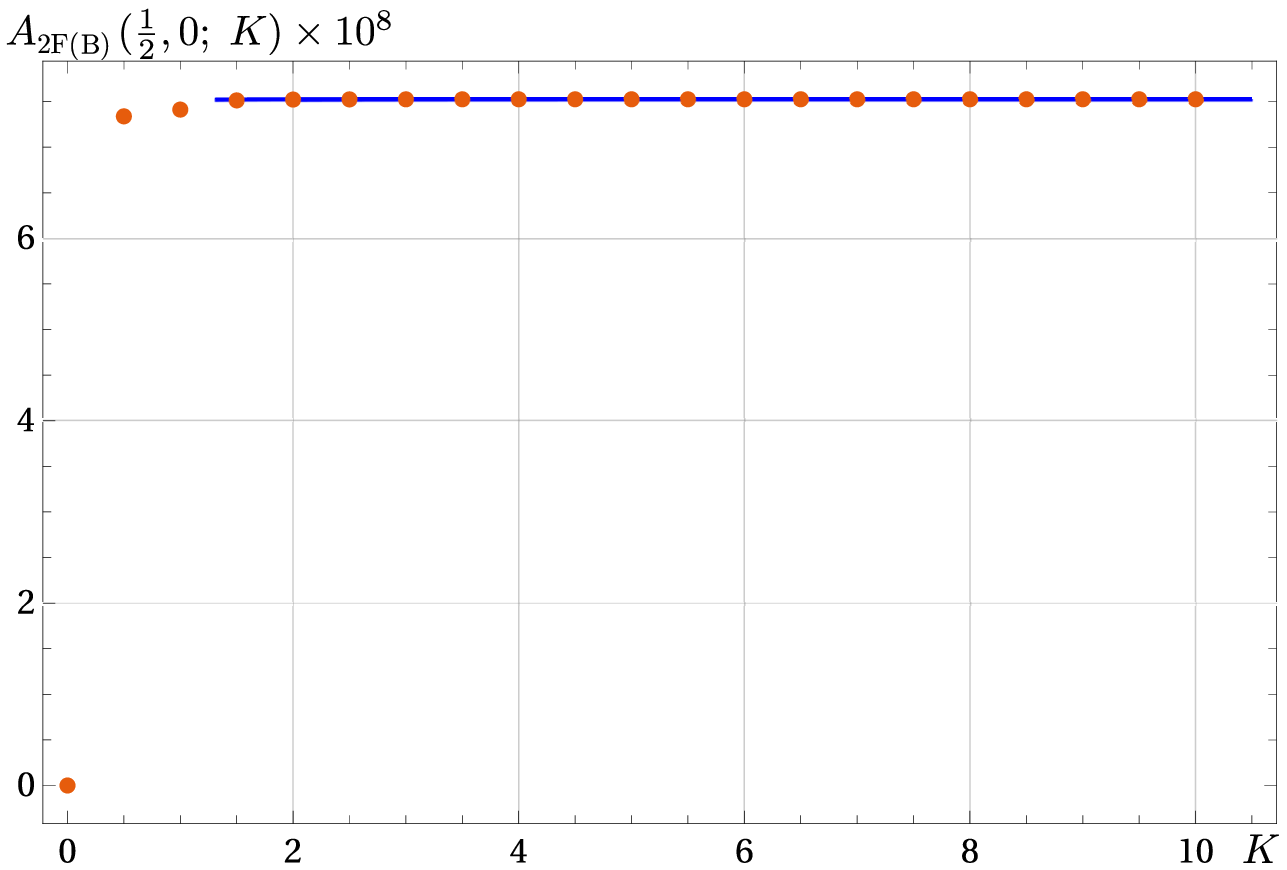}
\caption{\label{fig:2F_selfenergyplots}
Plot of the numerical values of the EPRL amplitudes with boundary spins $j_b=\tfrac{1}{2}$, boundary intertwiners $i_b=0$ and Immirzi parameter $\gamma=0.1$. We plot with a blue band the corresponding fits. 
Left panel: Spin foam diagram \dref{2F}{fig:2Fselfenergy1} and fit \eqref{eq:2F_selfenergyfinal1}.
Right panel: Spin foam diagram \dref{2F}{fig:2Fselfenergy2} and fit \eqref{eq:2F_selfenergyfinal2}.
}
\end{figure}

The amplitudes of the diagram \dref{2F}{fig:2Fselfenergy1} and \dref{2F}{fig:2Fselfenergy2} are 
\begin{equation}
    \label{eq:2F_selfenergy1}
    \begin{split}
     A_{2F(A)} \left(j_b, \  i_b; \, K\right) = \sum_{j_f = 0}^{K}  \sum_{i_e} & A_f(j_{f_1})A_f(j_{f_2}) \, A_e(i_{1})A_e(i_{2})A_e(i_{3})A_e(i_{4}) \times\\ 
     & A_v \left(  j_b, j_b, j_b, j_b, j_{f_1}, j_{b}, j_{f_2}, j_{f_1}, j_{b}, j_{f_1}; \ i_b , i_1, i_2, i_3, i_4\right) \times \\
     &  A_v \left(  j_b, j_b, j_b, j_b, j_{f_1}, j_{b}, j_{f_2}, j_{f_1}, j_{b}, j_{f_1}; \ i_b , i_4, i_2, i_3, i_1 \right) \ ,
    \end{split}
\end{equation}
and 
\begin{equation}
    \label{eq:2F_selfenergy2}
    \begin{split}
     A_{2F(B)} \left(j_b, \  i_b; \, K\right) = \sum_{j_f = 0}^{K}  \sum_{i_e} & A_f(j_{f_1})A_f(j_{f_2}) \, A_e(i_{1})A_e(i_{2})A_e(i_{3})A_e(i_{4}) \times\\ 
     & A_v \left(j_b, j_b, j_b, j_b, j_b, j_{f_1}, j_{f_1}, j_{f_2}, j_{f_1}, j_b; \ i_b , i_1, i_2, i_3, i_4\right) \times \\
     &  A_v \left(j_b, j_b, j_b, j_b, j_b, j_{f_1}, j_{f_1}, j_{f_2}, j_{f_1}, j_b; \ i_b , i_1, i_3, i_2, i_4\right) \ . 
    \end{split}
\end{equation}
Both amplitudes are convergent. With a fit with a $95\%$ confidence interval we find
\begin{equation}
\label{eq:2F_selfenergyfinal1}
 A_{2F(A)} (\tfrac{1}{2},0; K)  = (3.481, 3.482) \cdot 10^{-8} + (1.643,2.689)\cdot 10^{-10} \ K^{-1} - (6.924,5.082) \cdot 10^{-10} \ K^{-2} \ , 
\end{equation}
and 
\begin{equation}
\label{eq:2F_selfenergyfinal2}
 A_{2F(B)} (\tfrac{1}{2},0; K)  = (7.5237, 7.5243) \cdot 10^{-8} + (7.584,12.386)\cdot 10^{-11} \ K^{-1} - (3.245,2.400) \cdot 10^{-10} \ K^{-2} \ , 
\end{equation}
and we compare them with the numerical values of the amplitudes in Figure~\ref{fig:2F_selfenergyplots}.

Since these diagrams contain fewer faces than diagrams in Figure \ref{fig:4faces}, we expect them to converge more rapidly, and the plots showing the numerical results indeed suggest that this is the case. In an attempt to quantify the rate of convergence of these amplitudes, we amplify the divergence of the amplitudes by replacing the face amplitude in \eqref{eq:transitionamplitude} as
\begin{equation}
    A_f(j_f) = 2j_f + 1 \quad \longrightarrow \quad A_f^{amp}(j_f) = (2j_f + 1)^4 \ .
\end{equation}
With this amplification, we find a linearly divergent amplitude, suggesting that the original amplitude has a degree of divergence of $-5$.


\section{Conclusions}

In the absence of a cosmological constant, spin foam models are affected by infrared divergences. They are associated with large volume divergences and originate from the unbounded sums over the spin foam's bulk degrees of freedom.

\medskip
Calculating the radiative corrections to the Lorentzian EPRL spin foam propagator is necessary to study the theory's renormalization and explore the continuum limit. Two-vertex diagrams provide the first contributions in a vertex expansion. Previous studies focused on the `melonic' self-energy diagram \dref{6F}{fig:6Fselfenergy}, with numerical evidence suggesting it diverges linearly with respect to the spin cutoff \cite{Frisoni:2021uwx, Frisoni:2021dlk}, but there exist other diagrams with two vertices that also need to be computed to include all contributions at this order in the vertex expansion.

\medskip

In this paper, we studied two-vertex diagrams with two boundary edges, four boundary faces containing both boundary edges, and whose propagators preserve the order of the faces. We find eight additional diagrams that we classify based on the number of internal faces.

We focus on the diagrams with four and two bulk faces (see Figure~\ref{fig:4faces} and Figure~\ref{fig:2faces}). These diagrams are expected to be subdominant because they have fewer internal faces with respect to the diagram \dref{6F}{fig:6Fselfenergy}, and we find numerical evidence that, in fact, the diagrams with four or two internal faces appear to be convergent. We also point out the existence of a second diagram with six internal faces \dref{6F}{fig:6Fgoat}, but leave the numerical calculation of this diagram for future work since it requires computational resources we currently do not have access to.   

\medskip 

For simplicity, in the numerical calculations we fix the boundary spins to $\tfrac{1}{2}$, the first non-trivial option. We consider both possible boundary intertwiners ($0$ or $1$) and we repeat the calculations setting the Immirzi parameter to $\gamma = 0.1$ and also $\gamma = 1$ to minimize the dependence of our calculation on the specific choice of boundary data. 
We introduce a homogenous cutoff $K$ on the summations over the spins of the bulk faces, and we compute the EPRL spin foam amplitudes using the \texttt{sl2cfoam-next} framework as a function of $K$. For the diagrams with four or two internal faces, all amplitudes are found to have a degree of divergence (in $K$) smaller than the melonic self-energy diagram, confirming expectations. Somewhat surprisingly, these diagrams all appear to be convergent, with numerics suggesting a degree of divergence of $K^{-1}$ for diagrams with 4 internal faces, and an even faster convergence for diagrams with two internal faces. If this is the case then within the class of diagrams we consider, seven of the nine diagrams are finite (and of the two remaining diagrams, both have six internal faces, and one is known to diverge while the other has not been computed yet).

\medskip 

This result may have some important ramifications for renormalization. While we have only considered a certain class of diagrams, the fact that many are finite 
provides some encouragement in that the theory might be renormalizable. On the other hand, there seems to be a rapid growth in the number of diagrams, especially as the number of vertices is increased. Whether the spin foam model is renormalizable or not may hinge on which of these two effects dominates---of course, it will be necessary to consider diagrams with more vertices in order to answer this question and we leave this task for future work.

Related to this point is the question of which spin foam diagrams should be considered in a given calculation. In this paper, we imposed some conditions that reduced the number of diagrams that we considered, although it may be necessary to relax these conditions (for more on this point, see Appendix~\ref{app:other}). This question is closely related to the symmetries of the spin foam model: what exactly are the symmetries that should be imposed on a spin foam model, and do they force certain contributions to vanish?

It may be fruitful to consider the questions of renormalization and symmetries from the perspective of group field theory, whose partition function can be expanded in the usual way with the result giving a spin foam model \cite{Freidel:2005qe, Oriti:2013aqa}. It may be easier to impose symmetries directly in the group field theory action, and in turn this can be used to determine exactly which diagrams will contribute, simply by seeing whether they appear when calculating the Feynman expansion for the partition function. A further advantage is that this approach can also be used to calculate symmetry factors for different spin foam diagrams, something which may not be quite as easy to determine from the spin foam model on its own.

\medskip 

Finally, an important goal for future research would be to find a simple formula giving the degree of divergence of any diagram given its number of vertices, number of bulk edges, and number of bulk faces (and perhaps other relevant topological or combinatorial information); this could be useful for example to determine which interaction terms (in a group field theory language) are relevant/irrelevant. Our work suggests a clear dependence of the degree of divergence on the number of bulk faces, with more internal faces producing a greater degree of divergence. The melonic self-energy diagram appears to diverge linearly and has six bulk faces, while diagrams with four internal faces seem to have a degree of divergence $K^{-1}$, and diagrams with two internal faces converge even faster still. Nevertheless, it is clearly important to compute a wider range of diagrams to infer a general formula for the EPRL degree of divergence; we leave this for future work.


\section{Acknowledgments}


We thank Daniele Oriti for helpful discussions.
P.D.~is supported by the  FQXi  Grant  FQXi-RFP-1818 and of the ID\# 61466 grant from the John Templeton Foundation, as part of the ``The Quantum Information Structure of Spacetime (QISS)'' Project (\href{qiss.fr}{qiss.fr}).
P.F.~is supported by the Natural Science and Engineering Council of Canada (NSERC) through the Discovery Grant ``Loop Quantum Gravity: from Computation to Phenomenology''.
E.W.-E.~is supported by the Natural Sciences and Engineering Research Council of Canada, and by the UNB Fritz Grein Research Award.
We also acknowledge the Shared Hierarchical Academic Research Computing Network (SHARCNET) and Compute Canada (\href{https://www.computecanada.ca/}{www.computecanada.ca}) for granting access to their high-performance computing resources. 

We acknowledge the Anishinaabek, Haudenosaunee, L\=unaap\'eewak and Attawandaron peoples, on whose traditional lands Western University is located, and the Wolastoqiyik (Maliseet) peoples, on whose unsurrendered and unceded traditional lands the University of New Brunswick is located.

\begin{appendices}

\section{Details of the Lorentzian EPRL vertex amplitude}
\label{app:vertexdetails}

We report here the detailed definition of the  Lorentzian EPRL vertex amplitude.
\begin{equation}
    \begin{split}
    A_{v}\left(  j_1, j_2, j_3, j_4 , j_5, j_6, j_7,  j_8, j_9,  j_{10} ; \right. & \left. \ i_1 , i_2, i_3, i_4, i_5 \right) = \\
    \sum_{l_f=j_f}^{\infty}\sum_{k_e} &\left\lbrace \begin{array}{ccccc} 
    i_{1} & j_{3} & k_{4} & l_{6} & k_{2} \\
    j_{4} & l_{10} & l_{8} & l_{5} & j_{1} \\ 
    l_{7} & k_{5} & l_{9} & k_{3} & j_{2} 
    \end{array}\right\rbrace \\
  &(2k_2 + 1) (2k_3 + 1) (2k_4 + 1) (2k_5 + 1) \\
  &B_4^\gamma\left(j_{5} , j_{6}, j_{7}, j_{1}, l_{5}, l_{6}, l_{7}, j_{1}; i_{2},k_{2}\right) \\
  &B_4^\gamma\left(j_{8} , j_{9}, j_{2}, j_{5}, l_{8}, l_{9}, j_{2}, l_{5}; i_{3},k_{3}\right) \\
  &B_4^\gamma\left(j_{10} , j_{3}, j_{6}, j_{8}, l_{10}, j_{3}, l_{6}, l_{8}; i_{4},k_{4}\right) \\
  &B_4^\gamma\left(j_{4} , j_{7}, j_{9}, j_{10}, j_{4}, l_{7}, l_{9}, l_{10}; i_{5},k_{5}\right) \ .
\end{split}
\end{equation}
The $SU(2)$ invariant is a $\{15j\}$ symbol of the first kind and can be written in terms of Wigner's $\{6j\}$ symbols.
\begin{equation}
    \label{eq:15jsymbol}
    \begin{split}
        \left \{ \begin{array}{ccccc} j_1 & j_2 & j_3 & j_4 & j_5  \\  
                                      l_1 & l_2 & l_3 & l_4 & l_5  \\ 
                                      k_1 & k_2 & k_3 & k_4 & k_5 
                  \end{array}\right \} = 
  &(-1)^{\sum_{i=1}^5 j_i + l_i +k_i} \sum_x (2 x +1) \Wsix{j_1}{k_1}{x}{k_2}{j_2}{l_1} \Wsix{j_2}{k_2}{x}{k_3}{j_3}{l_2} \\ & \hspace{6mm }\times \Wsix{j_3}{k_3}{x}{k_4}{j_4}{l_3} \Wsix{j_4}{k_4}{x}{k_5}{j_5}{l_4} \Wsix{j_5}{k_5}{x}{j_1}{k_1}{l_5}  \ .
    \end{split}
\end{equation}      
The booster functions are a one dimensional integral over the rapidity $r$ of the reduce matrix elements in the $\gamma$-simple unitary representation of $SL(2,\mathrm{C})$.

\begin{equation}
\begin{split}
  \label{eq:boosterdef}
  B_4^\gamma&\left( j_1,j_2,j_3,j_4, l_1,l_2,l_3,l_4 ; i,k\right) 
  =\\
  &\sum_{ p_f } 
  \left(\begin{array}{cccc} l_1 & l_2 & l_3 & l_4 \\ p_1 & p_2 & p_3 & p_4 \end{array}\right)^{(k)}
  \left(\int_0^\infty \dd r \frac{1}{4\pi}\sinh^2r \, \bigotimes_{f=1}^4 d^{\gamma j_f,j_f}_{l_f j_f p_f}(r) \right)
  \left(\begin{array}{cccc} j_1 & j_2 & j_3 & j_4 \\ p_1 & p_2 & p_3 & p_4 \end{array}\right)^{(i)} 
  \ .
\end{split}
\end{equation}

The expression for $d_{jlm}^{\gamma j , j}(r)$ was given in \cite{Ruhl:1970fk, Speziale:2016axj}
\begin{equation}\label{eq:dSL2C}
\begin{split}
    d^{(\gamma j,j)}_{jlp}(r) =&  
    (-1)^{\frac{j-l}{2}} \frac{\Gamma\left( j + i \gamma j +1\right)}{\left|\Gamma\left(  j + i \gamma j +1\right)\right|} \frac{\Gamma\left( l - i \gamma j +1\right)}{\left|\Gamma\left(  l - i \gamma j +1\right)\right|} \frac{\sqrt{2j+1}\sqrt{2l+1}}{(j+l+1)!}  \\
    & \times \left[(2j)!(l+j)!(l-j)!\frac{(l+p)!(l-p)!}{(j+p)!(j-p)!}\right]^{1/2} e^{-(j-i\gamma j +p+1)r} \\
    & \times \sum_{s} \frac{(-1)^{s} \, e^{- 2 s r} }{s!(l-j-s)!} \, {}_2F_1[l+1-i\gamma j,j+p+1+s,j+l+2,1-e^{-2r}] \ .
\end{split}
\end{equation}
where ${}_{2}F_{1}$ is the Gauss hypergeometric function.

\section{Topological BF SU(2) spin foam theory}
\label{app:SU2}

Calculations with the topological BF SU(2) spin foam theory are straightforward analytically and numerically. Therefore they provide a convenient test for the techniques we use and the choice of parameters we make with the EPRL theory. 

The analytical calculation reduces to the integration of group valued delta functions and the numerical calculations can be comfortably done on a standard laptop using optimized libraries to compute Wigner symbols \cite{Wigxjpf_library, fastwigxj_related}.

The boundary data, the face, and the edge amplitudes of this theory are the same as the EPRL spin foam theory \eqref{eq:transitionamplitude}. The vertex amplitude is simpler and consists of a $SU(2)$ invariant $A_v(j_f,i_e) = \{15j\}$ \eqref{eq:15jsymbol}. The theory is affected by large spin divergences, and the renormalization is well studied \cite{Freidel:2002dw,Bonzom:2011br} and is related to residual diffeomorphism gauge symmetry.  We can rewrite the amplitude \eqref{eq:transitionamplitude} in a way that is more suitable for analytical calculations. The amplitude can be cast in terms of integrals (with the invariant measure) over a copy of $SU(2)$ for each edge of $SU(2)$ irreducible representations of spin $j_f$ for each face. Boundary edges are contracted with 4-valent intertwiners. Bulk (closed) faces the $SU(2)$ character $\Tr D^{(j_f)}$ of the product of all the group elements of the edges composing the face $g_f$. The sum over the spin associated with the face $j_f$ can be performed exactly
\begin{equation}
\label{eq:su2delta}
    \sum_{j_f} A_f(j_f) \Tr D^{(j_f)} (g_f) = \sum_{j_f} (2j_f+1) \Tr D^{(j_f)} (g_f) = \delta (g_f) \ ,
\end{equation}
in terms of the delta function over the group. The group elements are usually interpreted as the $SU(2)$ holonomy providing the parallel transport between reference frames associated to the spin foam edges. The delta function on spin foam faces is forcing holonomy around each bulk face to be trivial, as expected from a topological theory. In this form, the calculation of the amplitude reduces to integrating out delta functions. Divergences of the amplitudes manifest as delta functions evaluated at the identity. We regularize them with a cutoff $K$.

\begin{equation}
\label{eq:su2deltaK}
    \delta(\mathds{1}) \approx \sum_{j_f=0}^K (2j_f+1) \Tr D^{(j_f)} (\mathds{1}) = \sum_{j_f=0}^K (2j_f+1)^2 = \frac{8}{3} k^3 + 6 k^2+\frac{13}{3} k +1 =  O(K^3)\ .
\end{equation}

We compute the degree of divergence analytically of four diagrams in this paper \dref{6F}{fig:6Fselfenergy},  \dref{4F}{fig:4Ffrog1}, \dref{4F}{fig:4Fpenguin}, and  \dref{4F}{fig:4Fgoat1}.

\medskip
The BF amplitude of the melonic self-energy diagram \dref{6F}{fig:6Fselfenergy} is 
\begin{equation}
\begin{split}
    A^{(BF)}_{6F(A)} (j_b,i_b;\,K ) = \sum_{m_{1}m_{2}m_{3}m_{4}}&\left(\begin{array}{cccc}
j_{b} & j_{b} & j_{b} & j_{b}\\
m_{1} & m_{2} & m_{3} & m_{4}
\end{array}\right)^{i_{b}}\left(\begin{array}{cccc}
j_{b} & j_{b} & j_{b} & j_{b}\\
n_{1} & n_{2} & n_{3} & n_{4}
\end{array}\right)^{i_{b}}\\
\int \prod_{e=1}^6dg_{e} & D_{m_{1}n_{1}}^{j_{1}}\left(g_{1}g_{2}g_{6}\right)D_{m_{2}n_{2}}^{j_{2}}\left(g_{1}g_{3}g_{6}\right)D_{m_{3}n_{3}}^{j_{3}}\left(g_{1}g_{4}g_{6}\right)D_{m_{4}n_{4}}^{j_{4}}\left(g_{1}g_{5}g_{6}\right)\\
&\delta\left(g_{2}g_{3}^{-1}\right)\delta\left(g_{2}g_{4}^{-1}\right)\delta\left(g_{2}g_{5}^{-1}\right)\delta\left(g_{3}g_{4}^{-1}\right)\delta\left(g_{3}g_{5}^{-1}\right)\delta\left(g_{4}g_{5}^{-1}\right)\ . \\
\end{split}
\end{equation}
Performing all the integrals one by one, and using the invariance of the intertwiners we get
\begin{equation}
A^{(BF)}_{6F(A)} (j_b,i_b;\,K ) = \frac{1}{2i_b+1}\delta(\mathds{1})^3 = O(K^9) \ .
\end{equation}
The degree of divergence of the melonic self-energy diagram is $9$.

Similarly we can look at the BF amplitude of the diagram \dref{4F}{fig:4Ffrog1},
\begin{equation}
\begin{split}
    A^{(BF)}_{4F(A)} (j_b,i_b;\,K ) = \sum_{m_{1}m_{2}m_{3}m_{4}}&\left(\begin{array}{cccc}
j_{b} & j_{b} & j_{b} & j_{b}\\
m_{1} & m_{2} & m_{3} & m_{4}
\end{array}\right)^{i_{b}}\left(\begin{array}{cccc}
j_{b} & j_{b} & j_{b} & j_{b}\\
n_{1} & n_{2} & n_{3} & n_{4}
\end{array}\right)^{i_{b}}\\
\int \prod_{e=1}^6dg_{e} &  D_{m_{1}n_{1}}^{j_{b}}\left(g_{1}g_{2}g_{3}g_{4}g_{6}\right)D_{m_{2}n_{2}}^{j_{b}}\left(g_{1}g_{2}g_{3}g_{4}g_{6}\right)D_{m_{3}n_{3}}^{j_{b}}\left(g_{1}g_{2}g_{3}g_{4}g_{6}\right)D_{m_{4}n_{4}}^{j_{b}}\left(g_{1}g_{2}g_{3}g_{4}g_{6}\right)\\
& \delta\left(g_{2}\right)\delta\left(g_{4}\right)\delta\left(g_{3}g_{5}^{-1}\right)\delta\left(g_{2}g_{5}g_{4}g_{5}^{-1}\right)\ . \\
\end{split}
\end{equation}
The result of the integrals is
\begin{equation}
A^{(BF)}_{4F(A)} (j_b,i_b;\,K ) = \frac{1}{2i_b+1}\delta(\mathds{1}) = O(K^3) \ .
\end{equation}
The degree of divergence of the diagram \dref{4F}{fig:4Ffrog1} is $3$. Notice that the divergence of this diagram is subdominant to the melonic self-energy one.

In total analogy the BF amplitude of the diagram \dref{4F}{fig:4Fpenguin} is 
\begin{equation}
\begin{split}
    A^{(BF)}_{4F(C)} (j_b,i_b;\,K ) = \sum_{m_{1}m_{2}m_{3}m_{4}}&\left(\begin{array}{cccc}
j_{b} & j_{b} & j_{b} & j_{b}\\
m_{1} & m_{2} & m_{3} & m_{4}
\end{array}\right)^{i_{b}}\left(\begin{array}{cccc}
j_{b} & j_{b} & j_{b} & j_{b}\\
n_{1} & n_{2} & n_{3} & n_{4}
\end{array}\right)^{i_{b}}\\
\int \prod_{e=1}^6dg_{e} & D_{m_{1}n_{1}}^{j_{b}}\left(g_{1}g_{6}\right)D_{m_{2}n_{2}}^{j_{b}}\left(g_{1}g_{3}g_{5}g_{3}^{-1}g_{6}\right)D_{m_{2}n_{2}}^{j_{b}}\left(g_{1}g_{2}g_{5}g_{4}^{-1}g_{6}\right)D_{m_{2}n_{2}}^{j_{b}}\left(g_{1}g_{2}g_{5}^{-1}g_{4}^{-1}g_{6}\right)\\
&\delta\left(g_{2}g_{3}^{-1}\right)\delta\left(g_{2}g_{4}^{-1}\right)\delta\left(g_{3}g_{4}^{-1}\right)\delta\left(g_{5}\right)\ . \\
\end{split}
\end{equation}
The result of the integrals is
\begin{equation}
A^{(BF)}_{4F(C)} (j_b,i_b;\,K ) = \frac{1}{2i_b+1}\delta(\mathds{1}) = O(K^3) \ ,
\end{equation}
so the degree of divergence of the diagram \dref{4F}{fig:4Fpenguin}  is $3$.

Finally the BF amplitude of the diagram \dref{4F}{fig:4Fgoat1} is 
\begin{equation}
\begin{split}
    A^{(BF)}_{4F(D)} (j_b,i_b;\,K ) = \sum_{m_{1}m_{2}m_{3}m_{4}}&\left(\begin{array}{cccc}
j_{b} & j_{b} & j_{b} & j_{b}\\
m_{1} & m_{2} & m_{3} & m_{4}
\end{array}\right)^{i_{b}}\left(\begin{array}{cccc}
j_{b} & j_{b} & j_{b} & j_{b}\\
n_{1} & n_{2} & n_{3} & n_{4}
\end{array}\right)^{i_{b}}\\
\int \prod_{e=1}^6dg_{e} & \int D_{m_{1}n_{1}}^{j_{b}}\left(g_{1}g_{6}\right)D_{m_{2}n_{2}}^{j_{b}}\left(g_{1}g_{2}g_{6}\right)D_{m_{3}n_{3}}^{j_{b}}\left(g_{1}g_{2}^{-1}g_{6}\right)D_{m_{4}n_{4}}^{j_{b}}\left(g_{1}g_{3}g_{4}g_{5}g_{3}^{-1}g_{6}\right)\\
&\delta\left(g_{2}\right)\delta\left(g_{4}\right)\delta\left(g_{5}\right)\delta\left(g_{2}g_{3}g_{4}g_{5}g_{4}^{-1}g_{5}^{-1}g_{3}^{-1}\right)\ . \\
\end{split}
\end{equation}
The result of the integrals is
\begin{equation}
A^{(BF)}_{4F(D)} (j_b,i_b;\,K ) = \frac{1}{2i_b+1}\delta(\mathds{1}) = O(K^3) \ .
\end{equation}
Giving once again a degree of divergence of $3$ for the diagram \dref{4F}{fig:4Fgoat1}.

\medskip 

\begin{figure}[H]
\centering
        \includegraphics[width=0.495\textwidth]{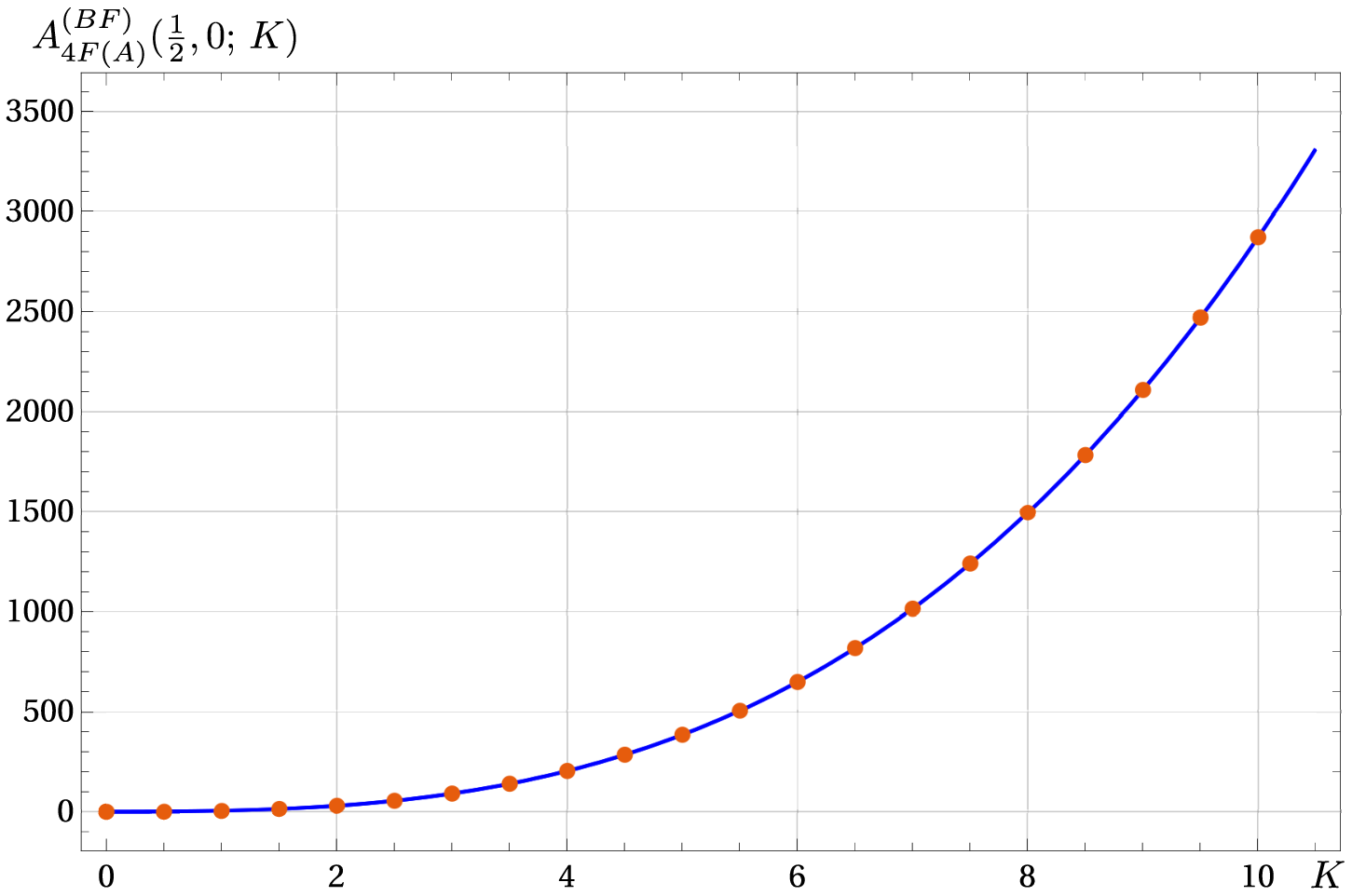}
        \includegraphics[width=0.495\textwidth]{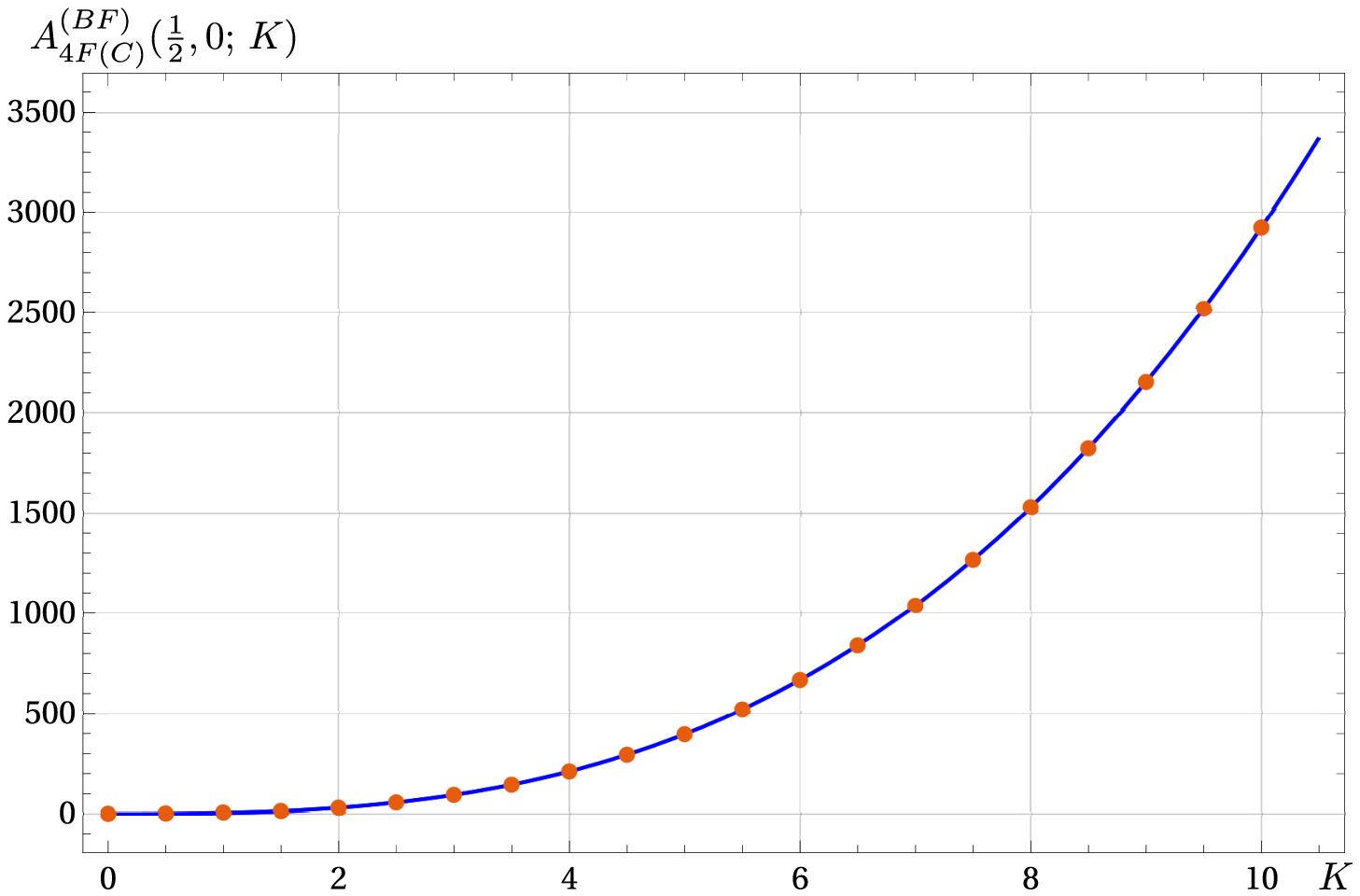}
        \includegraphics[width=0.495\textwidth]{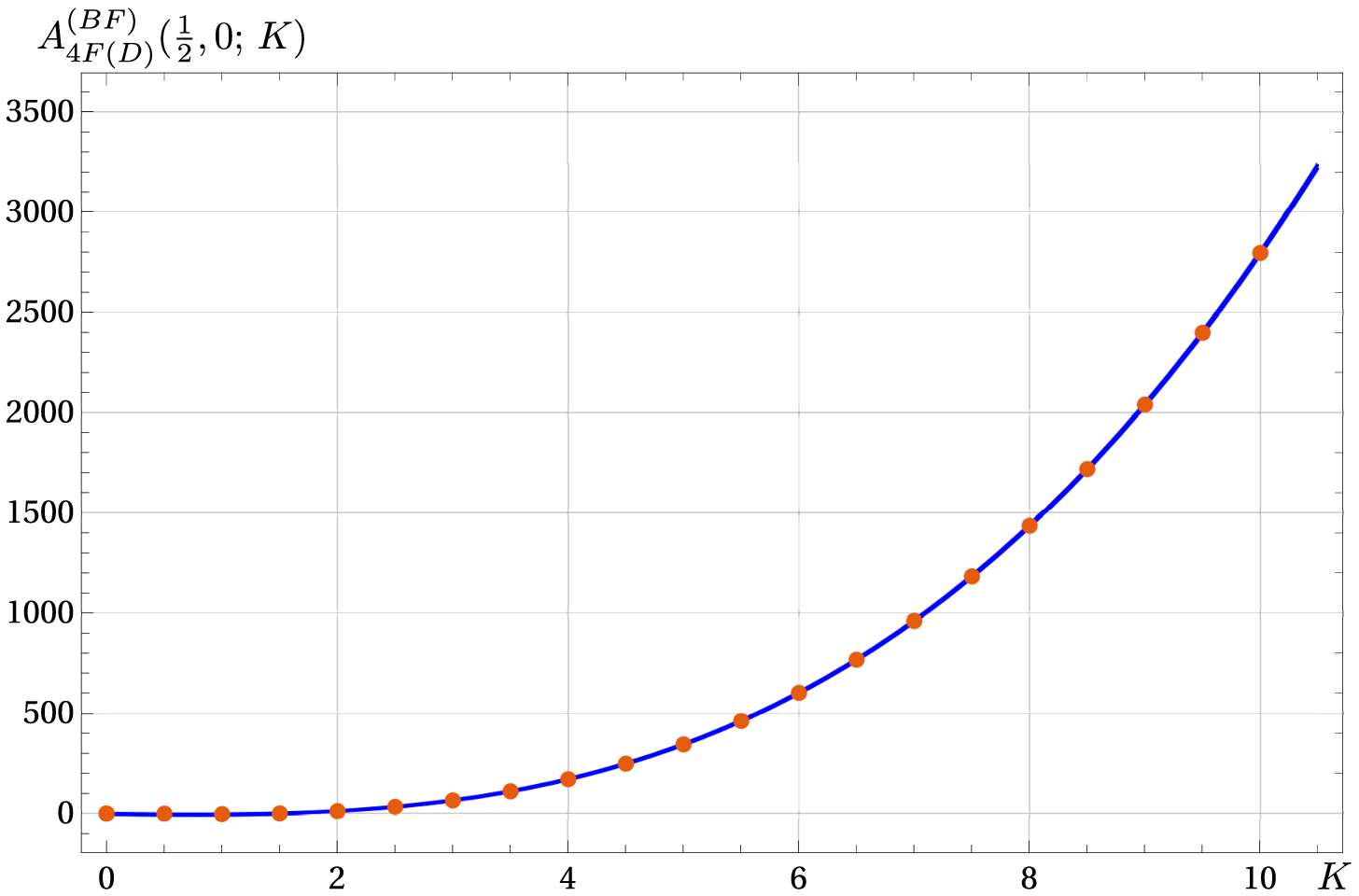}
\caption{\label{fig:BF} Plot of the numerical values of the BF amplitude of the spin foam diagram \dref{4F}{fig:4Ffrog1}, \dref{4F}{fig:4Fpenguin}, and  \dref{4F}{fig:4Fgoat1} (left to right) with boundary spins $j_b=\tfrac{1}{2}$ (orange dots) and the corresponding fits (blue band). The leading order of the fits are $A_{4F(A)}(\tfrac{1}{2},0; K) \approx (2.657,2.667)k^3$, $A_{4F(C)}(\tfrac{1}{2},0; K) \approx (2.666 , 2.667)k^3$, $A_{4F(D)}(\tfrac{1}{2},0; K) \approx (2.666, 2.667)k^3$. For all cases, it is compatible with the analytical value of $8/3 k^3$ of \eqref{eq:su2deltaK}.}
\end{figure}

We compare the analytic calculation of the degree of divergence of the diagrams \dref{4F}{fig:4Ffrog1}, \dref{4F}{fig:4Fpenguin}, and  \dref{4F}{fig:4Fgoat1} with the numerical estimation done using the same technique we employ for the EPRL model. The computation is straightforward and can be done in seconds using the $BF$ vertex of \texttt{sl2cfoam-next}. We fix the homogeneous cutoff to $K\lessapprox 10$ as a prototype for the EPRL calculation. In all three cases, we can fit the amplitudes with a cubic polynomial (see Figure~\ref{fig:BF}), finding a perfect agreement with the analytic calculations. Furthermore, it suggests that using a cutoff of order 10 is not a limiting factor in our analysis. 

\section{Other Diagrams with Two Vertices}
\label{app:other}

\begin{figure}
\centering
    \begin{subfigure}{0.49\textwidth}
        \centering
        \includegraphics[scale=0.25]{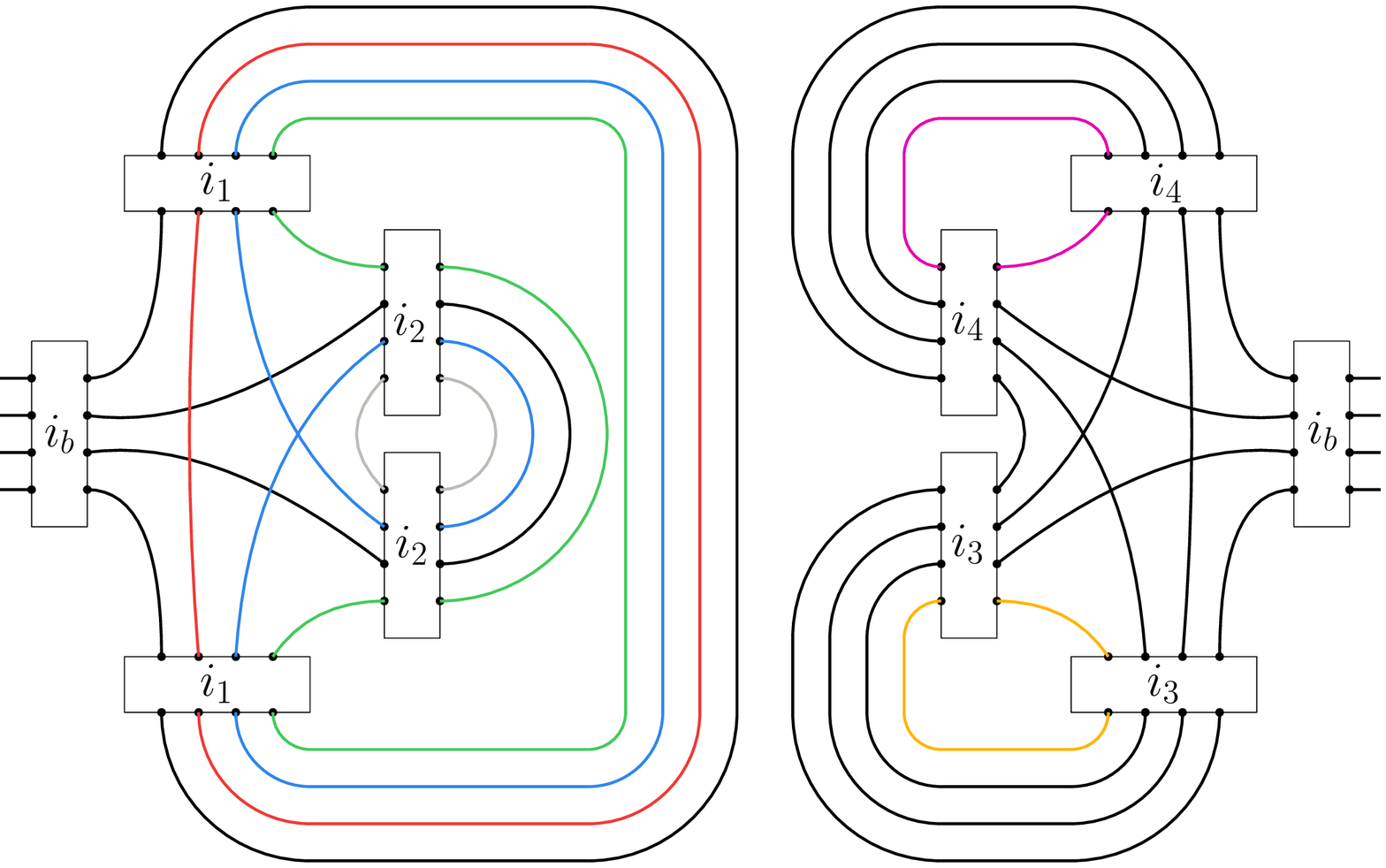}
        \caption{}
        \label{fig:Appendix1}
    \end{subfigure}
\hfill
    \begin{subfigure}{0.49\textwidth}
        \centering
        \includegraphics[scale=0.25]{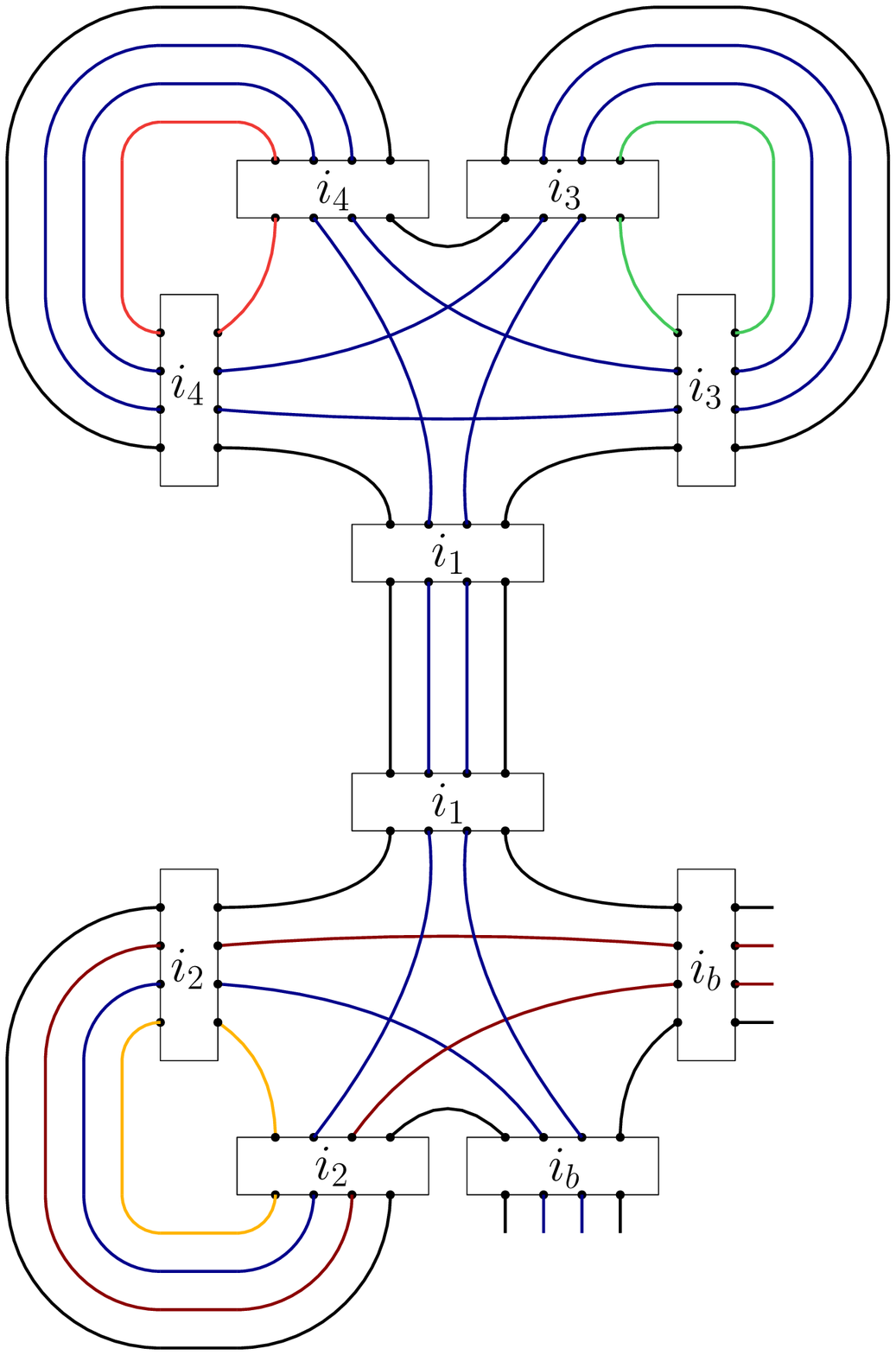}
        \caption{}
        \label{fig:Appendix2}
    \end{subfigure}
\hfill
    \begin{subfigure}{0.3\textwidth}
        \centering
        \includegraphics[scale=0.25]{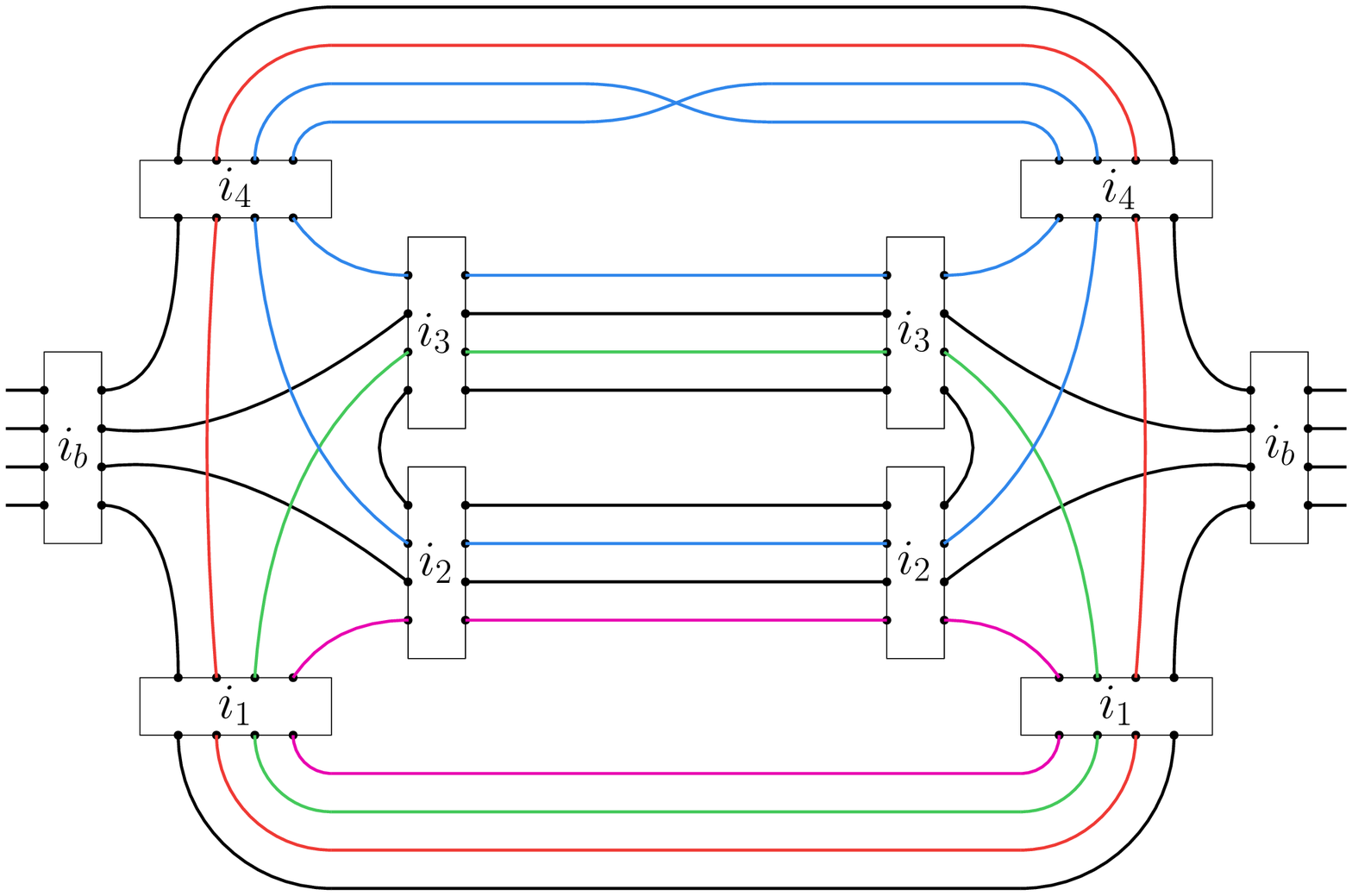}
        \caption{}
        \label{fig:Appendix3}
    \end{subfigure}
\hfill
    \begin{subfigure}{0.3\textwidth}
        \centering
        \includegraphics[scale=0.25]{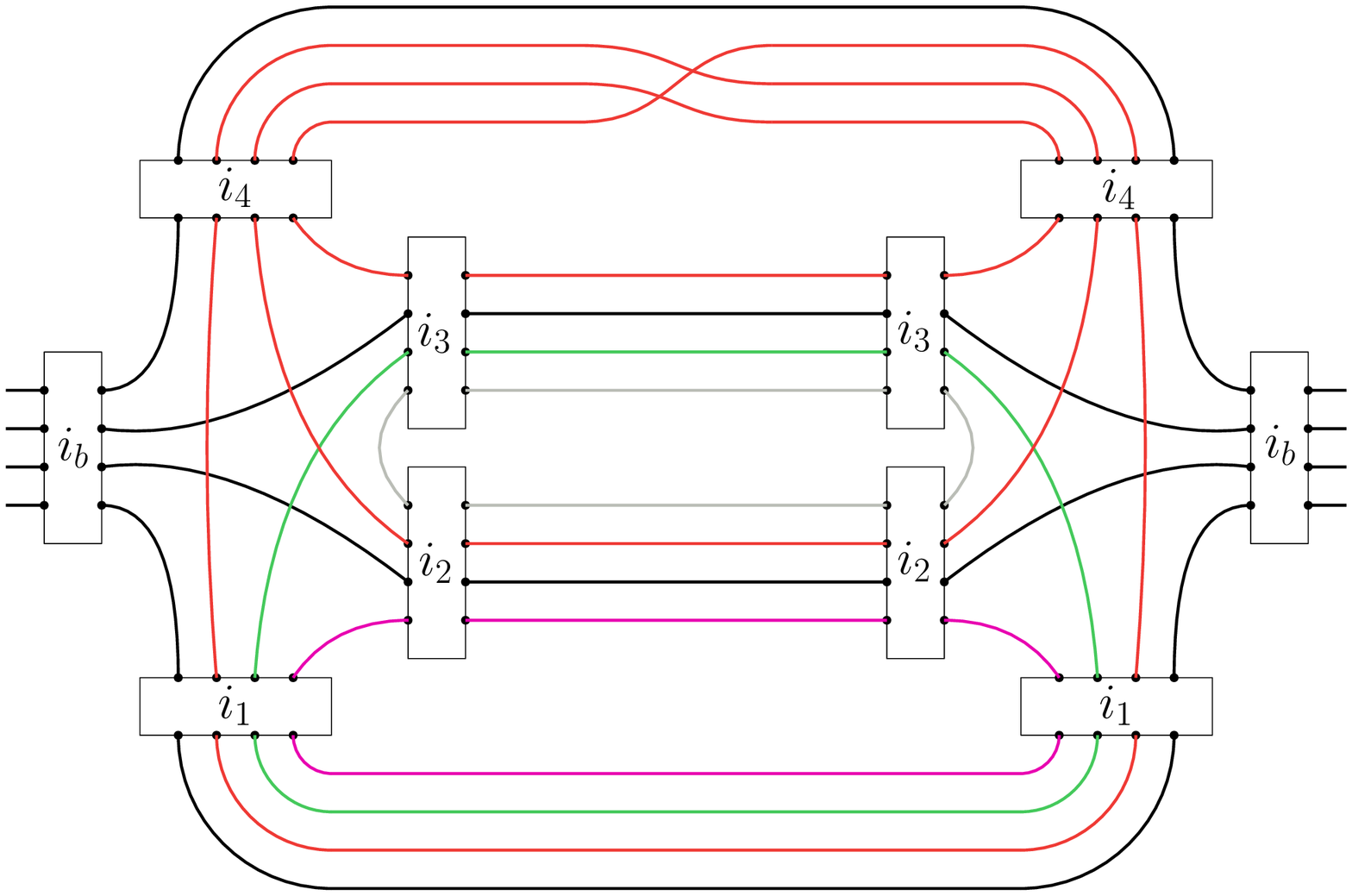}
        \caption{}
        \label{fig:Appendix4}
    \end{subfigure}
\hfill
    \begin{subfigure}{0.3\textwidth}
        \centering
        \includegraphics[scale=0.25]{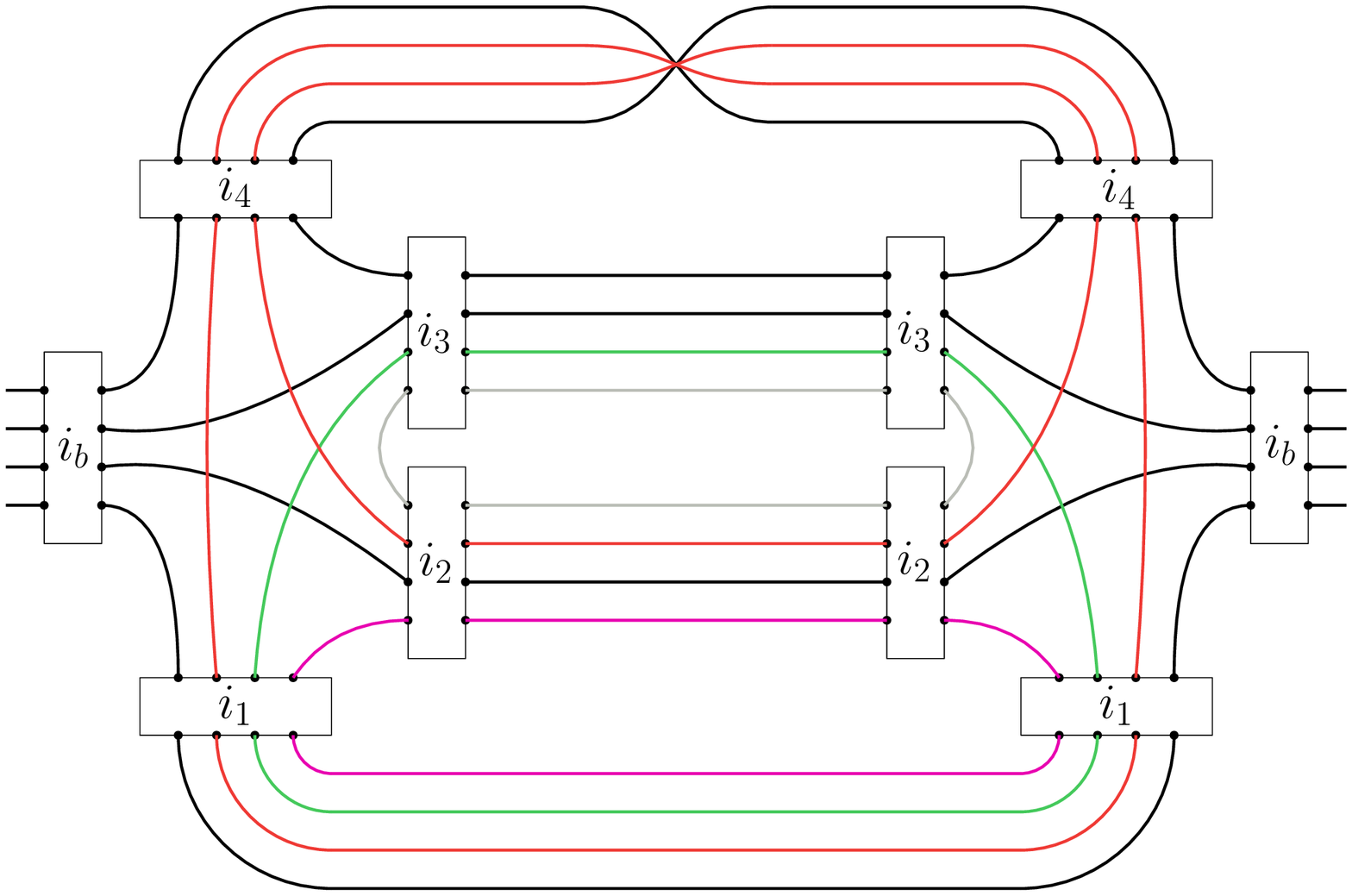}
        \caption{}
        \label{fig:Appendix5}
    \end{subfigure}
\caption{Some examples of spin foam diagrams we did not consider.}
\label{fig:other}
\end{figure}

Here we present some diagrams with two vertices and two boundary edges, but that are ruled out by (at least) one of the other two conditions that (i) there are exactly four boundary faces, and each boundary face contains both boundary edges, and (ii) the propagators are trivial and preserve the order of the faces. The examples we give here are not exhaustive (there are many more diagrams besides the ones we discuss here), and are only meant as a survey of other possible contributions to the self-energy calculation. Whether these diagrams (and others) should be included or not depends on the symmetries of the spin foam model. We note that we imposed these conditions as a simplifying assumption in order to reduce the number of diagrams we consider as a first step. To calculate the full self-energy of the propagator (to two vertices) it may be necessary to include the diagrams that do not satisfy these conditions.

\medskip

These two conditions are quite different. The first condition tracks the boundary faces through the spin foam and requires that each boundary face traverse the diagram and exit on the outgoing edge. Another way to understand this condition is that it requires all boundary faces to connect the initial and final boundary edges in the spin foam. This condition is not guaranteed to be satisfied because the boundary face could turn around and exit on the ingoing edge. A diagram with this property (a face that enters and leaves the diagram on the same edge) is only possible if two faces of the edge have the same spin, for example $j_1 = j_2$. For this reason, this type of diagram can only contribute to the self-energy calculation for specific types of edge propagators, namely those with faces that have at least one repeated spin. On the other hand, for an edge whose four faces all have different spins, these diagrams cannot contribute. Although it may seem unusual to have diagrams that contribute for some choices of boundary spins but not others (given the same boundary graph), if all possible paths are to be summed over there appear to be more paths possible in the presence of symmetries in the spin labels of the propagator.

In Fig.~\ref{fig:other}, the diagrams (A) and (B) are two examples of spin foams that violate condition (i). These two diagrams fall into two subcategories: diagram (A) is disconnected, while diagram (B) is connected. In diagram (A), the spin foam is completely disconnected, and it could contribute to the self-energy only for propagators where the four faces can be paired up, in this case $j_1=j_4$ and $j_2=j_3$. In contrast, diagram (B) is connected, and for this spin foam only two of the faces of the propagator need to be paired up, in this case $j_2=j_3$ (we highlight them in dark red and blue in the picture). 

As an aside, note that (either) half of diagram (A) may be relevant for no-boundary calculations in cosmology, where one wishes to calculate the transition from the `no-geometry vacuum' state (with no quanta of geometry) to an excited state corresponding to a cosmological space-time.

The second condition that the order of the faces in the propagator is preserved ensures that there are no `twists' in the internal propagators. We imposed this condition for simplicity because it drastically reduces the number of diagrams to consider. Taking the geometric perspective that the propagator represents a quantum tetrahedron, a cyclic permutation of three faces can be seen as a rotation, while the interchange of two faces is a reflection (or a parity transformation). Since there is no preferred background, it may seem natural that rotations should be a symmetry of the theory \cite{Ooguri:1992eb}, although there exist arguments that parity transformations should not be allowed \cite{Rovelli:2012yy, Christodoulou:2012af}. Note that any number of interchanges in the order of the faces corresponds to either a rotation or a parity transformation: an even number of interchanges of faces gives a rotation, while an odd number is a parity transformation (perhaps composed with a rotation).

Examples of these two possibilities are given in Fig.~\ref{fig:other}, where diagram (C) contains one interchange of faces corresponding to a parity transformation, while there is a cyclic permutation of three faces in diagram (D) corresponding to a rotation. Note that `twisted' diagrams will often have fewer internal faces than similar spin foams without any twists in the internal propagators. As a technical aside, note that the edge's intertwiner refers to a specific recoupling choice, so under a rotation or parity transformation it is important to track how the recoupling scheme transforms.

Also, note that a propagator that completely reverses the order of the faces from $(j_1, j_2, j_3, j_4)$ to $(j_4, j_3, j_2, j_1)$ can obtained through rotations and simply corresponds to a change in whether the propagator is being viewed from `above' or from `below'; an example is given in diagram (E) in Fig.~\ref{fig:other}. It may seem appropriate that this transformation be allowed for propagators in spin foams, even if other rotations and reflections are not.

Clearly, the spin foam diagrams that must be included in the calculation depend on the symmetries of the spin foam model. For example, the spin foam model derived from coloured group field theory has a very rigid combinatorics (thereby ensuring a one-to-one correspondence between any given spin foam diagram and the dual simplicial complex) that does not allow any permutations in the faces of the propagator \cite{Gurau:2009tw}. As a result, the number of spin foams that need to be computed (with a given number of vertices) for any given boundary state will be much smaller for coloured group field theories, thereby significantly simplifying the calculation. On the other hand, a model that allows permutations of the faces in the propagators will have many more diagrams contribute when calculating the self-energy, and presumably also when calculating other amplitudes as well.

To summarize, in this paper we imposed some conditions on the self-energy diagrams we considered in order to reduce the number of spin foams we need to calculate, as otherwise there is a large number of diagrams that need to be computed. We stress that these conditions are used only to simplify the task for a first exploration of self-energy diagrams beyond the melonic diagram shown in Figure~\dref{6F}{fig:6Fselfenergy} (that has already been studied in some detail \cite{Riello:2013bzw, Frisoni:2021uwx,Frisoni:2021dlk, Dona:2018pxq}). Although determining the correct conditions to be imposed on the self-energy diagrams will ultimately depend on the symmetries of the theory (these symmetries could be defined directly at the level of the spin foam model, or instead in a group field theory from which one can derive the spin foam expansion), it is possible that the conditions imposed here may be too strict, in which case it would be necessary to relax these conditions, depending on the symmetries of the spin foam model.

\end{appendices}


\providecommand{\href}[2]{#2}\begingroup\raggedright%

\endgroup

\end{document}